%% file: Defects.tex
\documentclass[a4paper, traditabstract,longauth]{aa}
\usepackage{graphicx,amsmath}
\usepackage{epsf}
\usepackage{color}
\usepackage[table,usenames,dvipsnames]{xcolor}

\usepackage{natbib}
\bibpunct{(}{)}{;}{a}{}{,} 

\input Planck.tex

\usepackage{txfonts}

\usepackage{fixltx2e}
\usepackage{txfonts}
\usepackage{url}
\usepackage{natbib}
\bibpunct{(}{)}{;}{a}{}{,}  
\newcommand\papertitle{\textit{Planck} 2013 results. XXV.  Searches for cosmic strings\\ and other topological defects}

\def\eqref#1{Eq.$\,$(\ref{#1})}

\newcommand{\eq}{\begin{eqnarray}}
\newcommand{\qe}{\end{eqnarray}}

\newcommand*\swallow[1]{}

\def\Gmu{G\mu/c^2}
\def\GmuAH{G\mu_{\rm AH}/c^2}

\newcommand{\boldmathsymbol}[1]{\ensuremath{\boldsymbol{#1}}}
\newcommand{\vect}[1]{\boldmathsymbol{#1}}

\newcommand{\uside}{\mathrm{side}}

\newcommand{\ud}{\mathrm{d}}

\newcommand{\healpix}{\texttt{HEALPix}}
\newcommand{\fwhm}{\mathrm{FWHM}}

\newcommand{\DTT}{\dfrac{\Delta T}{T}}
\newcommand{\Nside}{N_\uside}

\newcommand{\unitn}{\vect{\hat{n}}}
\newcommand{\Xp}{{X}'}
\newcommand{\Xd}{\dot{X}}
\newcommand{\ellmax}{\ell_{\max}}

\begin{document}
\title{\papertitle}
%

\input{AuthorList_P20_Strings_and_Other_Defects_authors_and_institutes}

\authorrunning{Planck Collaboration}
\titlerunning{\textit{Planck} 2013 results. XXV.  Searches for cosmic strings and other topological defects}

\bigskip
\bigskip
\bigskip

\abstract{\Planck\  data have been used to provide stringent new constraints on cosmic strings and other defects.  We describe forecasts of the CMB power spectrum induced by cosmic strings, calculating these from network models and simulations using line-of-sight Boltzmann solvers.   We have studied Nambu-Goto cosmic strings, as well as field theory strings for which radiative effects are important, thus spanning the range of theoretical uncertainty in the underlying strings models.  We have added the angular power spectrum from strings to that for a simple adiabatic model, with the extra fraction defined as $f_{10}$ at multipole $\ell=10$.  This parameter has been added to the standard six parameter fit using {\tt COSMOMC} with flat priors. For the Nambu-Goto string model, we have obtained a constraint on the string tension of 
$\Gmu < 1.5\times 10^{-7}$ and $ f_{10} < 0.015$ at 95\% confidence that can be improved to $\Gmu < 1.3\times 10^{-7}$ and $ f_{10} < 0.010$ on inclusion of high-$\ell$ CMB data.   
For the abelian-Higgs field theory model we find,
$\GmuAH < 3.2\times 10^{-7}$ and  $f_{10} < 0.028$.
The  marginalized likelihoods for $f_{10}$  and in the $f_{10}$--$\Omega_{\mathrm b}h^2$ plane are also presented. We have additionally obtained comparable constraints on $f_{10}$ for models with semilocal strings and global textures.  In terms of the effective defect energy scale these are somewhat weaker at $G \mu/c^2 < 1.1 \times 10^{-6}$.   We have made complementarity searches for the specific non-Gaussian signatures of cosmic strings, calibrating with all-sky \Planck\ resolution  CMB maps 
generated from networks of post-recombination strings.  We have validated our non-Gaussian searches using these simulated maps in a \Planck-realistic context, estimating sensitivities of up to $\Delta\Gmu \approx 4 \times 10^{-7}$.   
We have obtained upper limits on the string tension at 95\% confidence of
$\Gmu < 8.8\times 10^{-7}$ with modal bispectrum estimation and $\Gmu < 7.8\times 10^{-7}$ for real space searches with Minkowski functionals.    These are conservative upper bounds because only post-recombination string contributions have been included in the non-Gaussian analysis.}


\titlerunning{Cosmic strings and other topological defects}

\maketitle
\clearpage
%
%
\section{Introduction}

This paper, one of a set associated with the 2013 release of data from the \Planck\footnote{\Planck\ (\url{http://www.esa.int/Planck}) is a project of the European Space Agency (ESA) with instruments provided by two scientific consortia funded by ESA member states (in particular the lead countries France and Italy), with contributions from NASA (USA) and telescope reflectors provided by a collaboration between ESA and a scientific consortium led and funded by Denmark.} mission \citep{planck2013-p01}, describes the constraints on cosmic strings, semi-local strings and global textures.   Such cosmic defects 
are a generic outcome of symmetry-breaking phase transitions in the early Universe \citep{Kibble:1976sj} and further motivation 
came from a potential role in large-scale structure formation \citep{Zeldovich:1980gh,Vilenkin:1981iu}.  Cosmic
strings appear in a variety of supersymmetric and other grand unified
theories, forming at the end of
inflation~(see, for example, \citealt{Jeannerot:2003qv}). However, further interest in
cosmic (super-)strings has been motivated by their emergence in
higher-dimensional theories for the origin of our Universe, such as
brane inflation. These superstring variants come in a number of
$D-$ and $F-$string forms, creating hybrid networks with more complex
dynamics (see, e.g., \citealt{Polchinski:2004hb}).  Cosmic strings can have an enormous energy per unit length
$\mu$ that can give rise to a number of observable effects, including
gravitational lensing and a background of gravitational waves.  Here,
we shall concentrate on the impact of strings on the cosmic microwave
background (CMB), which includes the generation of line-like discontinuities in
 temperature.  Comparable effects can also be caused by other 
 types of cosmic defects, notably semi-local strings
and global textures.  As well as influencing the CMB power spectrum,
each type of topological defect should have a counterpart non-Gaussian
signature giving us the ability distinguish between different defects,
alternative scenarios, or systematic effects.  The discovery of any of
these objects would profoundly influence our understanding of
fundamental physics, identifying GUT-scale symmetry breaking patterns,
perhaps even providing direct evidence for extra dimensions.
Conversely, the absence of these objects will tightly constrain
symmetry breaking schemes, again providing guidance for high energy
theory.  For a general introduction to  cosmic strings and other defects, 
refer to \citet{vilenkin2000cosmic, Hindmarsh:1994re, Copeland:2009ga}.

High resolution numerical simulations of cosmic strings using the
Nambu-Goto action indicate that cosmological networks tend towards a
scale-invariant solution with typically tens of long strings stretching
across each horizon volume. These strings continuously source
gravitational perturbations on sub-horizon scales, the magnitude of
that are determined by the dimensionless parameter: \eq
\frac{G\mu}{c^2} = \left( \frac{\eta}{m_{\rm Pl}}\right)^2\,, \qe
where $\eta$ is the energy scale of the string-forming phase
transition and $m_{\rm Pl}\equiv \sqrt{\hbar c/G}$ is the \Planck\ mass.
String effects on the CMB power spectrum have been estimated using a
phenomenological string model and, with \textit{WMAP} and SDSS data, these
estimates yield a $2\sigma$ upper bound of $\Gmu < 2.6\times 10^{-7}$
\citep{Battye:2010xz}.  A consequence is that strings can be
responsible for no more than 4.4\% of the CMB anisotropy signal at
multipole $\ell=10$.

As we shall discuss,  the evolution of Nambu-Goto string networks is computationally challenging and quantitative uncertainties remain, notably in characterizing the string small-scale structure and loop production.  An alternative approach has been to use field theory simulations of cosmological vortex-strings. These yield a significantly lower number of strings per horizon volume (less than half), reflecting the importance of radiative effects on the microphysical scales being probed numerically.   The degree of convergence with Nambu-Goto string simulations is difficult to determine computationally at present, but there are also global strings for which radiative effects of comparable magnitude are expected to remain important on cosmological scales.   It is  prudent in this paper, therefore, to constrain both varieties of strings, labelling the field theory constraints as ${\rm AH}$ from the abelian-Higgs (local $U(1)$) model used to describe them.   Given these quantitative differences, such as the lower density, field theory strings produce a weaker constraint $\GmuAH < 5.7\times 10^{-7}$ using \textit{WMAP} data alone  \citep{Bevis:2007gh} and \citep{Urrestilla:2011gr}.   The shape of the string-induced power spectrum also has a different shape, which allows up to a 9.5\% contribution at $\ell =10$.     These \textit{WMAP} constraints can be improved by adding small-scale CMB anisotropy in a joint analysis.  The Nambu-Goto strings limit improves to become $\Gmu < 1.7\times 10^{-7}$ (using SPT data, \citealt{Dvorkin:2011aj}) and field theory strings yield $\GmuAH < 4.2\times 10^{-7}$ \citep{Urrestilla:2011gr}.  Power-spectrum based constraints on global textures were studied in \citet{Bevis:2004wk} and \citep{Urrestilla:2007sf}, with the latter paper giving a 95\% limit of $\Gmu<4.5\times 10^{-6}$. \citet{Urrestilla:2007sf} also provide constraints on semi-local strings, $\Gmu < 5.3 \times 10^{-6}$. 

Constraints on cosmic strings from non-Gaussianity require high resolution realisations of string-induced CMB maps that are extremely challenging to produce.  Low resolution small-angle and full-sky CMB maps calculated with the full recombination physics included, have indicated some evidence for a significant kurtosis from strings \citep{Landriau:2010cb}.  More progress has been made increating high resolution maps from string lensing after recombination (see \citet{Ringeval:2012tk} and references therein) and identifying, in principle, the bispectrum and trispectrum, which can be predicted for strings analytically \citep{Hindmarsh:2009qk, Hindmarsh:2009es,Regan:2009hv}.   The first \textit{WMAP} constraint on cosmic strings using the analytic CMB trispectrum yielded $\Gmu < 1.1\times 10^{-6}$ at 95\% confidence \citep{Fergusson:2010gn}. An alternative approach is to fit pixel-space templates to a map, this method was applied to global textures templates in \citet{Cruz:2007pe} and \citet{Feeney:2012hj,Feeney:2012jf}.

The most  stringent constraints that are claimed for the string tension arise from predicted backgrounds of gravitational waves that are created by decaying loops \citep{Vilenkin:1981bx}.  However, these constraints are strongly dependent on uncertain string physics, most notably the network loop production scale and the nature of string radiation from cusps, i.e., points on the strings approaching the speed of light $c$.  The most optimistic constraint based on the European Pulsar Timing Array is $\Gmu < 4.0\times 10^{-9}$ \citep{vanHaasteren:2011ni}, but a much more conservative estimate of  $\Gmu < 5.3\times 10^{-7}$ can be found in \citet{Sanidas:2012ee}, together with a string parameter constraint survey and an extensive discussion of these uncertainties.   Such gravitational wave limits do not apply to global strings or to strings for which other radiative channels are available.   

Alternative topological defects scenarios also have strong motivations and we report limits on textures and global monopoles in this paper as well. Of particular recent interest are hybrid networks of cosmic strings where the correlation length is reduced by having several interacting varieties (e.g., $F$- and $D$-strings) or a small reconnection probability, $p<1$. We expect to investigate these models using the \Planck\ full mission data. 

The outline of this paper is as follows.  In Sect.~2 we briefly describe the different types of topological defects that we consider, and their impact on the CMB anisotropies. We also discuss how the CMB power spectrum is computed and how we obtain CMB maps with a cosmic string contribution. In Sect.~3 we present the defect constraints from the CMB power spectrum (with numbers given in Table \ref{tab:limits}), while Sect.~4 discusses searches for topological defects with the help of their non-Gaussian signature. We finally present the overall conclusions in Sect.~5.

\section{Theoretical Modelling and Forecasting}

\subsection{Cosmic strings and their cosmological consequences}

\subsubsection{String network evolution}

A detailed quantitative understanding of the cosmological evolution of string networks is an essential pre-requisite for making accurate predictions about the cosmological consequences of strings.   Fortunately, all string network simulations to date have demonstrated convincingly that the large-scale properties of strings approach a self-similar scale-invariant regime soon after formation.  If we treat the string as a one-dimensional object, then it sweeps out a two-dimensional worldsheet in spacetime
\eq
x^\mu = x^\mu(\zeta^a), \qquad a= 0,\,1,
\qe
where the worldsheet parameters $\zeta^0$ and $\zeta^1$ are time-like and space-like respectively.    The Nambu-Goto action that governs string motion then becomes 
\eq \label{eq:Nambu}
S = \,- \,\mu\kern-2pt \int \kern-4pt\sqrt{-\gamma}\,\, d^{\,2}\zeta ,
\qe
where $\gamma_{ab} = g_{\mu\nu} \partial_a x^\mu \partial_b x^\nu$ is the two-dimensional worldsheet metric ($\gamma = \det(\gamma_{ab})$) induced by the spacetime metric  $ g_{\mu\nu}$.     The Nambu-Goto action \eqref{eq:Nambu} can be derived systematically from a field theory action, such as that for the abelian-Higgs model describing $U(1)$ vortex-strings:
\begin{equation}
\begin{aligned}
S =  \int d^4x \sqrt{- g} \,\left[(D_\mu\phi)^* (D^\mu\phi) - \frac{1}{4e^2} F_{\mu\nu} F^{\mu\nu}  - \frac{\lambda}{4} (|\phi|^2 - \eta^2)^2\right], 
\end{aligned}
\label{eq:aHiggs}
\end{equation}
where $\phi$ is a complex scalar field, $F^{\mu\nu}$ is the $U(1)$
field strength and $D_\mu = \partial_\mu + i A_\mu$ is the
gauge-covariant derivative with $e$ and $\lambda$ dimensionless
coupling constants.  The transverse degrees of freedom in $\phi$ can
be integrated out provided the string is not strongly curved, that is,
the string width $\delta \approx \hbar c/\eta \ll L$ where $L$ is the
typical radius of curvature.  For a cosmological string network today
with $\Gmu\sim 10^{-7}$, these two lengthscales are separated by over
40 orders of magnitude, so this should be a valid approximation.   

In an expanding universe, the Nambu-Goto action \eqref{eq:Nambu} yields a Hubble-damped wave equation governing the string motion.   These equations can be solved numerically, provided ``kinks'' or velocity discontinuities are treated carefully. However, they can also be averaged analytically to describe the scale-invariant evolution of the whole string network in terms of two quantities, the energy density $\rho$ and the r.m.s. velocity $v$.    
   Any string network divides fairly neatly into two distinct populations of long (or ``infinite'') strings $\rho_\infty$ stretching beyond the Hubble radius and the small loops $\rho_l$ with length $l \ll H^{-1}$ that the long strings create \cite{Kibble:1984hp}.   Assuming the long strings form a Brownian random walk characterised by a correlation length $L$, we have 
\eq\label{eq:onescale}
 \rho_\infty  = \frac{\mu}{L^2} ~\equiv~  \frac{\zeta\mu}{t^2}\,,
\qe
and the averaged equations of motion become simply
\eq
2\frac{dL}{dt}&=&2HL(1 + {v^2})
+{\tilde c}v \, , \label{eq:vos}\cr
\frac{dv_\infty}{dt}&=&\left(1-{v^2}\right)
\left[\frac{k(v)}{L}-2Hv\right]\, ,
\qe 
where $\tilde c $ measures the network loop production rate and  $k(v)$ is a curvature parameter with $k\approx 2\sqrt2(1- \sqrt2v)$.  This is the velocity-dependent one-scale (VOS) model and, with a single parameter $\tilde c$, it provides a good fit to both Nambu and field theory simulations, notably through the radiation-matter transition \citep{Martins:1996jp}.

A general consensus has emerged from the three main simulation codes
describing Nambu-Goto string networks
\citep{Martins:2005es,Ringeval:2005kr,BlancoPillado:2011dq}.  These
independent codes essentially solve for left- and right-moving modes
along the string using special techniques to handle contact
discontinuities or kinks, including ``shock fronting", artificial
compression methods and an exact solver for piecewise linear strings,
respectively.  The consistency between simulations is shown in
Table~\ref{tab:stringdensity} for the string density parameter $\zeta$
defined in \eqref{eq:onescale}.  Averaging yields the radiation era
density $\zeta =10.7$ and a matter era value $\zeta =3.3$.  Note that
these asymptotic values and the intervening matter-radiation
transition can be well-described by the VOS model \eqref{eq:vos} with
$\tilde c = 0.23$.  The matter era VOS value appears somewhat
anomalous from the other two simulations, but this is obtained from
larger simulations in a regime where convergence is very slow, so it
may more closely reflect the true asymptotic value.  These simulations
have also advanced the study of string small-scale structure and the loop
distribution, about which there had been less consensus (see, e.g., \citealt{BlancoPillado:2011dq}).  However, note that
CMB anisotropy is far less sensitive to this issue compared to
constraints from gravitational waves.

\begin{table}[tmb]
\begingroup
\newdimen\tblskip \tblskip=5pt
\caption{Summary of numerical simulation results for the string density parameter $\zeta$ defined in \eqref{eq:onescale}.  The Nambu-Goto string simulations are respectively labelled as  MS \citep{Martins:2005es}, RSB \citep{Ringeval:2005kr}, and BOS \citep{BlancoPillado:2011dq}. This is contrasted with the much lower density results from lattice field theory simulations of vortex-strings labelled as MMS \citep{Moore:2001px} and BHKU \citep{Bevis:2006mj}.}
\label{tab:stringdensity}
\nointerlineskip
\vskip -3mm
\footnotesize
\setbox\tablebox=\vbox{
   \newdimen\digitwidth 
   \setbox0=\hbox{\rm 0} 
   \digitwidth=\wd0 
   \catcode`*=\active 
   \def*{\kern\digitwidth}
   \newdimen\signwidth 
   \setbox0=\hbox{+} 
   \signwidth=\wd0 
   \catcode`!=\active 
   \def!{\kern\signwidth}

         \halign{\hbox to 1.15in{#\leaderfil}\tabskip 1.0em&
            \hfil#\hfil&
            \hfil#\hfil&
            \hfil#\hfil&
            \hfil#\hfil&
            \hfil#\hfil\tabskip 0pt\cr
    \noalign{\doubleline\vskip 2pt}
  Epoch & MS				  	  &  RSB  		&  BOS    & MSM  & BHKU\cr
   \noalign{\vskip 3pt\hrule\vskip 4pt}
  Radiation&11.5& 9.5&11.0&5.0&3.8\cr
Matter&3.0&3.2&3.7&1.5&1.3\cr
 \noalign{\vskip 3pt\hrule\vskip 4pt}
}}
\endPlancktablewide

\endgroup
\end{table}

Field theory simulations using lattice gauge techniques have also been employed to study the evolution of string networks in an expanding universe.   Comparatively, these three-dimensional  simulations are constrained to a lower dynamic range and the simulations require the solution of modified field equations to prevent the string core width shrinking below the lattice resolution. On the other hand, field theory simulations include field radiation and therefore provide a more complete account of the string physics. In Table~\ref{tab:stringdensity} the lower string densities obtained from two sets of abelian-Higgs simulations are given \citep{Moore:2001px,Bevis:2006mj}.  The evolution can be fitted with a VOS model  \eqref{eq:vos} with $\tilde c = 0.57$, which is 150\% higher than for Nambu-Goto strings.  Field theory simulations have further important applications, particularly for describing delocalised topological defects such as textures, for describing models that do not form stable defects like semilocal strings, and because they include radiative effects naturally.  Radiative effects observed in current abelian-Higgs simulations are comparable to the radiative damping anticipated for cosmological global strings and so the AH analysis below should offer some insight into this case. 

\begin{figure}
\begin{center}
\includegraphics[width=3.0in]{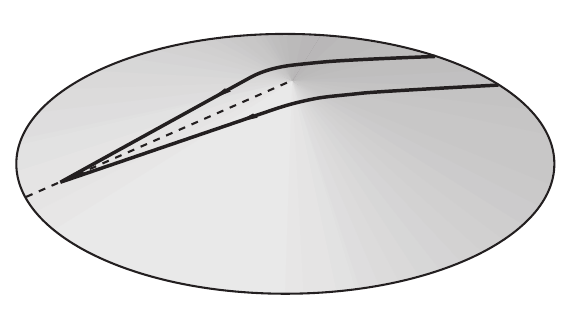}
\end{center}
\caption{The spacetime around a cosmic string is conical, as if a 
narrow wedge were removed from a flat sheet and the edges identified.
For this reason cosmic strings  
can create double images of distant objects. Strings moving across the line of sight will cause line-like discontinuities in the CMB radiation. }
\label{fig:grav_lens}
\end{figure}

\subsubsection{String gravity and the CMB}

Despite the enormous energy per unit length $\mu$, the spacetime around a straight cosmic string is locally flat.   The string
has an equation of state $p_z = - \rho$, $p_x=p_y=0$ (for one lying along the
$z$-direction), so there is no source term in the relativistic version of the 
Poisson equation $\nabla^2\Phi = 4\pi G(\rho + p_x+p_y+p_z)$.  The
straight string exhibits no analogue of the Newtonian pull of gravity on 
any surrounding matter.  But this does not mean the string has no gravitational
impact; on the contrary, a moving string has dramatic effects on
nearby matter or propagating CMB photons.  

The spacetime metric about a straight static string takes the simple form,
\eq
ds^2 = dt^2 -dz^2 -dr^2 -r^2d\theta^2\,,
\qe
that looks like Minkowski space in cylindrical coordinates, but for the fact
that the azimuthal coordinate $\theta$ has a restricted range $0\le \theta\le 
2\pi (1-4G\mu)$.  The spacetime is actually {\it conical} with a
global deficit angle
$\Delta = 8\pi G\mu$, that is, 
an angular wedge of width $\Delta$ is
removed from the space and the remaining edges identified (see Fig.~\ref{fig:grav_lens}).
This means that distant galaxies on the opposite side of a cosmic string can be gravitationally lensed to produce characteristic double images. 

\begin{figure}
\begin{center}
\includegraphics[width=0.75\hsize]{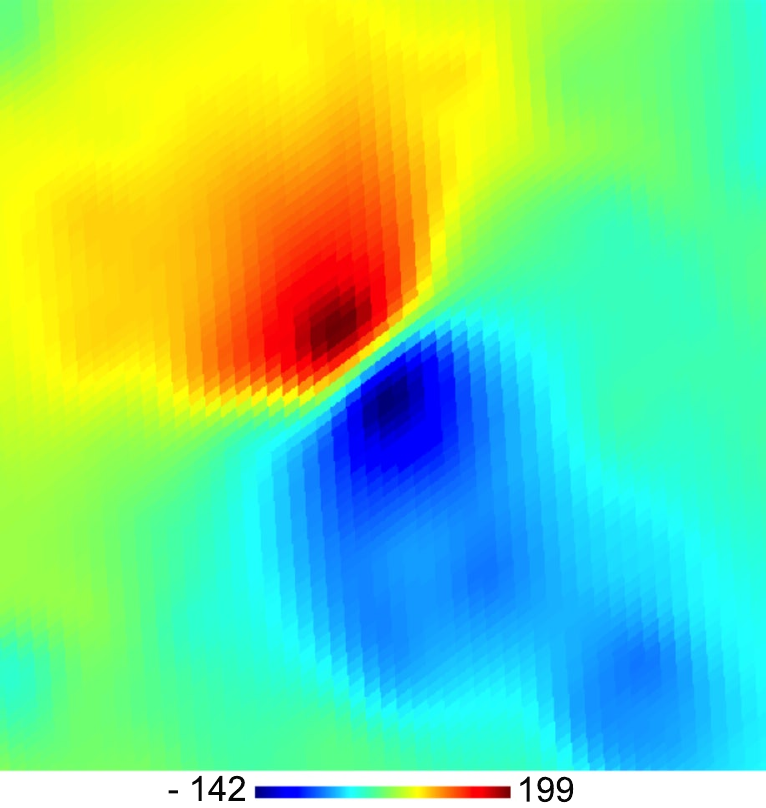}
\end{center}
\caption{Characteristic CMB temperature discontinuity created by a cosmic string.  Here, the simulated Nambu-Goto string has produced a  cusp, a small region on the string that approaches the speed of light, which has generated a  localised CMB signal.}
\label{fig:string_cusp}
\end{figure}

Cosmic strings create line-like discontinuities in the CMB signal. As
the string moves across the line of sight, the CMB photons are boosted
towards the observer, causing a relative CMB temperature shift across
the string, given by~\citep{Gott:1984ef,Kaiser:1984}  
\eq
\label{eq:cmbstrings}
\frac{\delta T}{T} = 8\pi G\mu v_{\rm s}\,\gamma_{\rm s}\,,
\qe
where $v_{\rm s}$ is the transverse velocity of the string and $\gamma_{\rm s} =
(1-v_{\rm s}^2)^{-1/2}$. This rather simple picture, however, is
complicated in an expanding universe with a wiggly string network and
relativistic matter and radiation components.  The energy-momentum
tensor $T_{\mu\nu}({\bf x}, t)$ essentially acts as a source term for the metric
fluctuations that perturb the CMB photons and create temperature
anisotropies.  Essentially, the problem can be recast using Green's (or transfer)
functions $G^{\mu\nu}$ that project forward the contributions of
strings from early times to today: \eq\label{eq:dtotstrg} \frac{\Delta
  T}{T}(\hat {\bf n}, {\bf x}_{\rm obs}, t_0) = \int d^4 x\, G^{\mu\nu}
(\hat {\bf n}, {\bf x}, {\bf x}_{\rm obs}, t, t_0)\,T_{\mu\nu}({\bf x}, t)\,,
\qe 
where $\hat{\bf n}$ is the line-of-sight direction for photon propagation and ${\bf x}_{\rm obs}$ is the observer position.
The actual quantitative solution of this problem entails a
sophisticated formalism to solve the Boltzmann equation and then to
follow photon propagation along the observer's line-of-sight.  An
example of the line-like discontinuity signal created by a cosmic
string in the CMB is shown in Fig.~\ref{fig:string_cusp}.  In this
case, a string cusp has formed on the string, causing a strongly
localised signal and reflecting the Lorentz boost factor in
\eqref{eq:cmbstrings}.

\subsection{Semi-local strings}

The tight constraints on the presence of cosmic strings that we will discuss later in this paper start to put pressure on the wide class of inflation models that generate such defects \citep{Hindmarsh:2011qj}. The power of these constraints would be reduced if the strings could be made unstable. This is the basic motivation behind semilocal strings: a duplication of the complex scalar field $\phi$ in the abelian-Higgs action (\ref{eq:aHiggs}), occurring naturally in a range of inflation models \citep{Dasgupta:2004dw,Achucarro:2005vz,Dasgupta:2007ds}, transforms the stable cosmic strings into non-topological semilocal strings \citep{Vachaspati:1991dz} as the vacuum manifold becomes $S^3$, which is simply-connected. The existence and stability of the semilocal strings is thus a question of dynamics rather than due to the topology of the vacuum manifold. In general we do not expect to form long strings, but rather shorter string segments, as the semilocal strings can have ends. The evolution of these segments is very complicated and arises directly from the field evolution, so that it is only practicable to simulate these defects with the help of field theory \citep{Urrestilla:2007sf}.

\subsection{Global defects}

A large alternative class of defects is due to the breaking of a global $O(N)$ symmetry (rather than a gauge symmetry as in the case of cosmic strings) of a $N$-component scalar field $\phi$. The energy density of global defects is significantly less localised than those that result from gauge symmetry breaking  due to the absence of the screening by a gauge field, and there are thus long-range forces between the defects. The field self-ordering is therefore very efficient for all types of defects with $N\geq2$, leading to a generic scaling of the defect energy density with the background energy density (see e.g.,\citealt{Durrer:2001cg}). For this reason global monopoles ($N=3$) do not overclose the Universe as their local counterparts would. In this paper we study specifically the case $N=4$ called ``texture'', which can arise naturally in many multi-field inflation models that involve a non-zero vacuum expectation value and symmetry breaking. In this case there are no stable topological defects present, but contrary to local texture, global texture can have a non-negligible impact on the perturbations in the cosmos, with the field self-ordering leading to ``unwinding events''. In spite of their non-topological nature, the field evolution is  closely related to the one of lower-dimensional stable global defects due to the long-range nature of the forces. This is similar to the case of the non-topological semilocal strings of the previous section, and indeed the semilocal example can be seen as an intermediate case between cosmic strings and global texture: Starting from the semilocal action, we can on the one hand revert to the cosmic string action by removing one of the complex scalar fields, and on the other hand we find the texture action if we remove the gauge field.

The normalization of global defects is usually given in terms of the parameter $\varepsilon = 8 \pi G \eta^2/c^2$ when using an action like Eq.\ (\ref{eq:aHiggs}) (with a second complex scalar field but without the gauge fields). However, for a simpler comparison with the cosmic string results we can recast this in terms of $\Gmu\equiv \varepsilon/4$ and quote limits on $\Gmu$ also for the texture model, as in  \citet{Urrestilla:2007sf}.

\subsection{CMB power spectra from cosmic defects}

The CMB power spectrum from topological defects, like strings, is more difficult to compute than the equivalent for inflationary scenarios that predict a spectrum dominated by an adiabatic component with a possible, but highly constrained, isocurvature component. In defect-based scenarios the perturbations are sourced continuously throughout the history of the Universe, as opposed to adiabatic and isocurvature modes that are the result of initial conditions. In principle this requires knowledge of the source, quantified by the unequal-time correlator (UETC) of the defect stress-energy tensor, from the time of defect formation near the GUT scale to the present day---a dynamic range of about $10^{52}$---something that will never be possible to simulate. Fortunately, we can use the scaling assumption to extrapolate the results of simulations with substantially smaller dynamic range. This has allowed a qualitative picture to emerge of the characteristics of the power spectra from defects, though quantitative predictions differ. Here, we will focus on spectra calculated in two different ways for cosmic strings, as well as spectra from semilocal strings and texture models.

Defect-based power spectra are dominated by different physical effects across the range of angular scales. (i) On large angular scales the spectra are dominated by an integrated Sachs-Wolfe (ISW) component due to the strings along the line-of-sight between the time of last scattering and the present day. The scaling assumption implies that this component will be close to scale invariant, although in practice it typically has a mildly blue spectrum. (ii) At intermediate scale the dominant contribution comes from anisotropies created at the time of last scattering. In contrast to the strong series of acoustic peaks created in adiabatic and isocurvature models, defects produce only a broad peak because their contributions are not coherent. (iii) At very small angular scales, the spectra are again dominated by the ISW effect because, rather than decaying exponentially due to the effects of Silk damping, there is only power-law decay with the exponent being a characteristic of the specific type of defect.

The standard lore is to treat the defect stress energy tensor, $\theta_{\mu\nu}$, as being covariantly conserved at first order, which is known as the ``stiff approximation''. In principle, this means that it is necessary to measure two independent quantities from the simulations, or model them. The other two component are then computed from the conservation equations. In practice things are a little more complicated since it is necessary to provide the UETC
\begin{equation}
U_{\mu\nu\alpha\beta}(k,\tau,\tau^{\prime})=\langle\theta_{\mu\nu}(k,\tau)\theta_{\alpha\beta}(k,\tau^{\prime})\rangle\,,
\end{equation}
where $\tau$ is the conformal time and $k$ is the wavenumber. Once one has the UETC, then there two ways to proceed. The first involves creating realisations of the defect stress-energy whose power spectra are computed then averaged to give the total power spectrum. The other approach involves diagonalization of the UETC.
During pure matter or radiation domination, the scaling property of defect evolution means that quantities are measured relative to the horizon scale, so that the
UETC is only a function of $x = k \tau$ and $x' = k \tau'$. These functions $U(x,x')$ can be discretized and then are symmetric matrices that we can diagonalize. The resulting eigenvectors can be inserted as sources into a Boltzmann code, and the resulting $C_\ell$ are then summed up, weighted by the eigenvalues \citep{Pen:1997ae,Durrer:2001cg}. Even though the power spectrum resulting from each ``eigen-source'' exhibits a series of acoustic peaks, the summation over many such spectra smears them out, as they are not coherent (unlike inflationary perturbations). This smearing-out explains why defect power spectra generically are smooth, as mentioned above.

\begin{figure*}[t]
\sidecaption
\centering
\includegraphics[width=110mm]{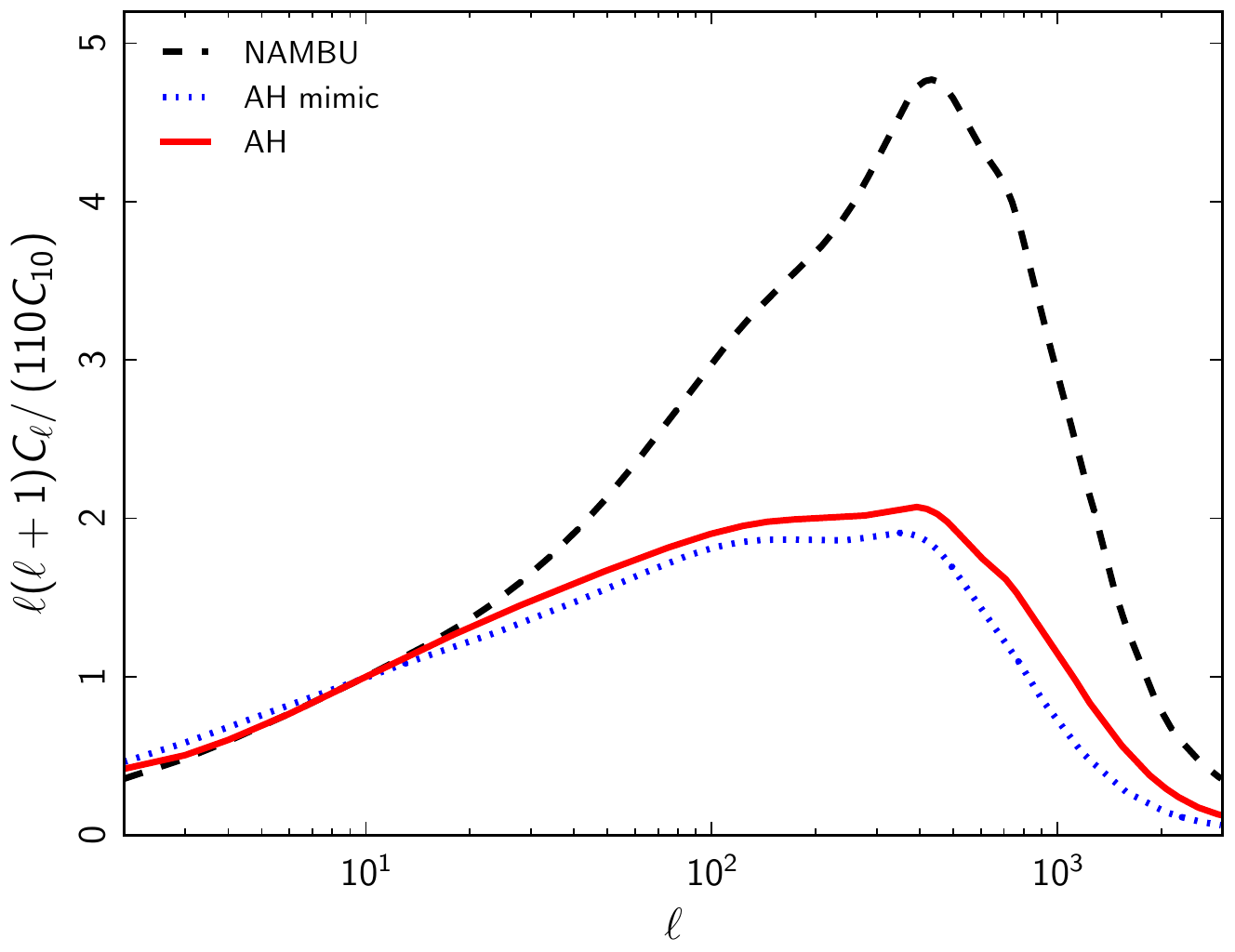}
\caption{Cosmic string power spectra used in this analysis: NAMBU (black dashed), AH-mimic (blue dotted) and AH (red solid). 
The spectra have been normalized to equal power at $\ell=10$.
The spectra are normalized the observed \textit{WMAP}7 value at $\ell=10$ and have $\Gmu=1.17\times 10^{-6}$, $1.89\times 10^{-6}$ and $2.04\times 10^{-6}$ respectively. 
Note that the limits discussed in this paper mean that the CMB spectra presented here are less than $3\%$ of the overall power spectrum amplitude and hence the differences observed at high $\ell$ do not have a much effect.\bigskip\bigskip} 
\label{fig:comp}
\end{figure*}

\begin{figure*}
\sidecaption
\centering
\includegraphics[width=110mm]{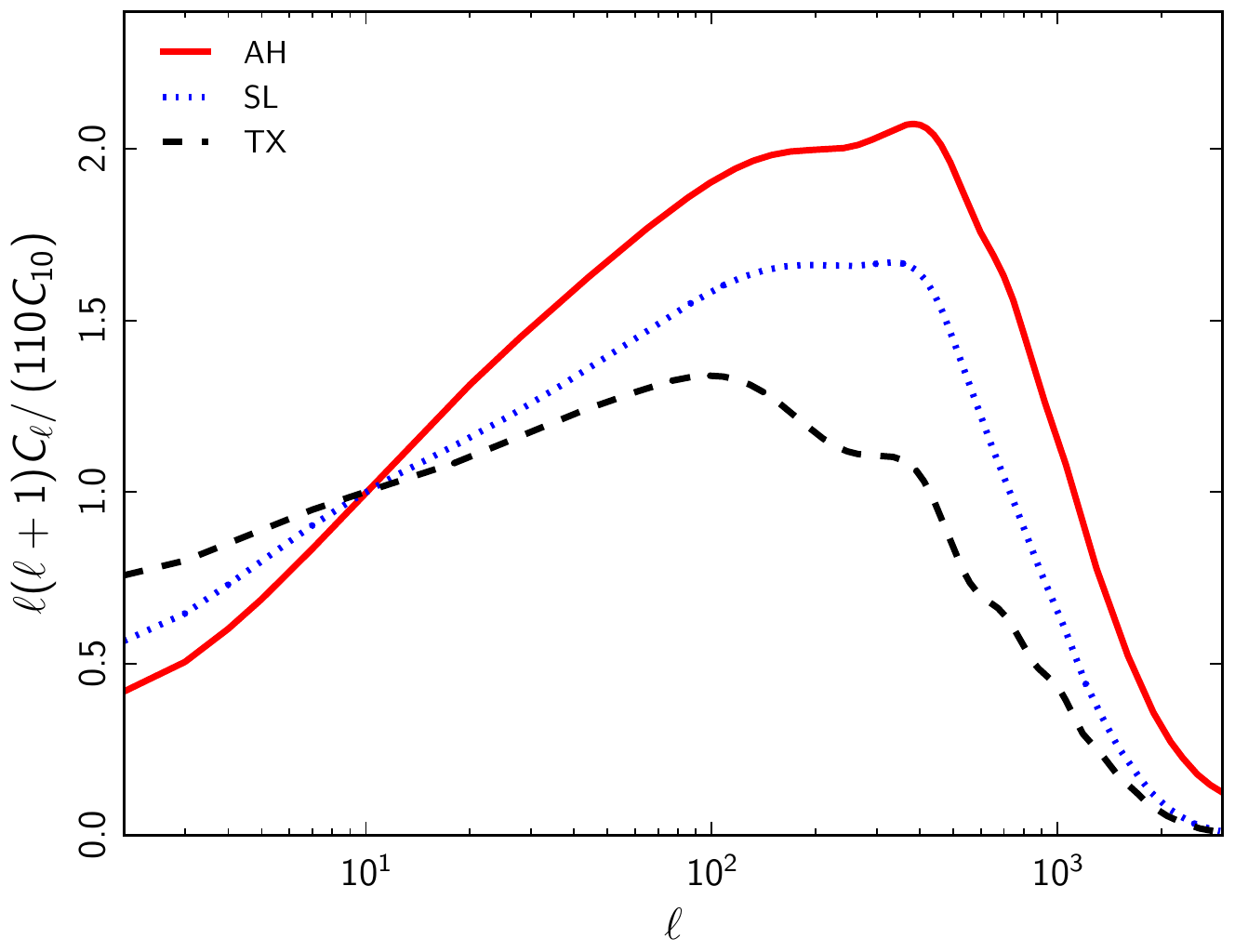}
\caption{Comparison between global texture (black dashed) and semilocal (blue dotted) string power spectra and the AH field theory strings (red solid), normalized to unity at $\ell=10$. As expected, the SL spectrum lies in between the TX and the AH spectra. The AH spectrum was recomputed for the \Planck\ cosmological model with sources from \cite{Bevis:2010gj}, and the SL and TX spectra were taken from \cite{Urrestilla:2007sf}.\bigskip\bigskip}
\label{fig:textures}
\end{figure*}

There are also several methods to obtain predictions for the UETCs of cosmic strings and other topological defects.
The first approach we will consider for cosmic strings is to use what has become known as the Unconnected Segment Model (USM; \citealt{Albrecht:1997nt,Albrecht:1997mz,Pogosian:1999np}). In its simplest form this models the cosmic string energy momentum tensor as that of an ensemble of line segments of correlation length $\xi d_{\rm H}(t)$, moving with an r.m.s.\ velocity $\langle v^2\rangle^{1/2}$, where $d_{\rm H}(t)$ is the horizon distance. In addition one can take into account the effects of string `wiggles" due to small-scale structure via a coefficient, $\beta=\mu_{\rm eff}/\mu$ quantifying the ratio of the renormalized mass per unit length to the true value. The model parameters $\xi$, $\langle v^2\rangle^{1/2}$ and $\beta$ are computed from simulations.   In our calculations we link the USM sources to the line-of-sight Boltzmann solver  \texttt{CMBACT} \citep{Pogosian:1999np} to create an ensemble of realisations from which we find an averaged angular power spectrum.   

There are two USM-based models that we will use in this analysis which we believe span the realistic possibilities---we note a more general approach marginalizing over three string parameters  is proposed in \citet{2011PhRvD..84d3522F} (see also recent work in \citealt{Avgoustidis:2012gb}). The first USM model, which we will refer to as {NAMBU}, is designed to model the observational consequences of simulations of cosmic string simulations performed in the Nambu-Goto approximation. In these simulations the scaling regime is different in the radiation and matter eras, with $(\xi,\langle v^2\rangle^{1/2}/c,\beta)_{\rm rad}=(0.13,0.65,1.9)$ and $(\xi,\langle v^2\rangle^{1/2}/c,\beta)_{\rm mat}=(0.21,0.60,1.5)$ and the extrapolation between the two is modelled by using the velocity dependent one-scale model \citep{Martins:1996jp}. In the second, which we will refer to as  AH-mimic, we attempt to model the field theory simulations using the Abelian-Higgs model described below, with $(\xi,\langle v^2\rangle^{1/2}/c,\beta)=(0.3,0.5,1)$ independent of time.

The other approach that we will consider is to measure the UETC directly from a simulation of cosmic strings in the Abelian-Higgs model, which we will refer to as AH. The Abelian Higgs model involves a complex scalar field $\phi$ and a gauge field $A_\mu$ described earlier \eqref{eq:aHiggs}, for which the 
dimensionless coupling constants $e$ and $\lambda$ are chosen with $\lambda = 2 e^2$, so that the characteristic scales 
of the magnetic and scalar energies are equal, (see \citet{Bevis:2006mj,Bevis:2007qz} for further details about the
model). We then simulate the evolution of the fields on a grid, starting from random initial conditions designed to
mimic a phase transition, followed by a brief period of diffusive evolution, to rapidly reach a scaling solution
expected to be typical of the configuration found long after the phase transition. As the simulation is
performed in comoving coordinates, the string width is effectively decreasing as time passes. To enlarge the dynamical
range available, we partially compensate this shrinking with an artificial string fattening. We perform runs for
various values of the fattening parameter to ensure that the results are not affected by it.

During the simulations, we compute the energy-momentum tensor at regular intervals and decompose it into scalar, vector and tensor parts.
We store these components once scaling is reached, and compute the UETCs by correlating them with later values of
the energy-momentum tensor. UETCs from several runs are averaged, diagonalized and then fed into a modified version of the \texttt{CMBEASY} Boltzmann code 
\citep{Doran:2003sy} to compute the CMB power spectra (both temperature and polarization). The spectra used in 
this paper were derived from field-theory simulations on a $1024^3$ grid and used the extrapolation to sub-string scales
described in \citet{Bevis:2010gj}, which are expected to be accurate at the 10\% level to $\ell_{\rm max} \approx 4000$.

In Figs.~\ref{fig:comp} and \ref{fig:textures} we present the spectra we will use in subsequent analysis. The higher dashed black curve is the spectrum computed using the USM for the {NAMBU} model, and the smaller dashed blue and solid red curves the AH-mimic model and the AH model, respectively. We should note that when normalized to the amplitude of the observed CMB anisotropies on large-scales at $\ell=10$, the three models give $\Gmu=1.17\times 10^{-6}$, $1.89\times 10^{-6}$ and $1.9\times 10^{-6}$ for the {NAMBU}, AH-mimic and AH models, respectively. The reasons for differences between the spectra for these two approaches are discussed in \citet{Battye:2010xz}. Briefly, the main reasons for the differences are twofold: First, the overall normalization, which is due to the NAMBU models having smaller values of $\xi$, more strings per horizon volume, and larger values of $\beta$, with each of the string segments being heavier, than the two AH models. Both these effects mean that a lower value of $\Gmu$ is required to achieve the same amplitude for the anisotropies. Secondly, the enhanced peak at small angular scales, which is caused by the value of $\xi$ being smaller in the radiation era than in the matter era, meaning that there are more strings per horizon volume in the radiation era when the small-scale anisotropy is imprinted, and hence more anisotropy on those scales for a given $\Gmu$.

The method used for the semilocal strings (denoted SL) and $O(4)$ global texture (denoted TX) is fundamentally the same as for the AH model: we simulate the field theory on a discretized grid and compute the energy-momentum tensor at regular intervals. From these snapshots we derive the UETCs by correlating the scalar, vector and tensor parts at different times. The only difference is the field-theory action being used in the simulations. In Fig.~\ref{fig:textures} we present the spectra we used for the semilocal strings and global textures, taken from \cite{Urrestilla:2007sf}. These models are also shown with the AH cosmic string model for comparison.

\subsection{Maps of CMB anisotropies from cosmic strings}

In order to go further than the two-point correlation function, we
have used numerical simulations of Nambu-Goto cosmic string evolution
in an FLRW spacetime to generate various CMB synthetic maps. The use of
simulations is crucial to produce realistic string configurations on
our past light cone and have been the subject of various code
development in the last twenty years (see \citealt{Albrecht:1989,
  Bennett:1989, Bennett:1990, Allen:1990,
  Vincent:1998,Moore:2002,Ringeval:2005kr,
  BlancoPillado:2011dq}). Until recently, the underlying numerical
challenges have limited the resolution of the full sky maps to an
angular resolution of $14'$ (corresponding to a $\healpix$ resolution
of $\Nside=256$) in \citet{Landriau:2002fx, 
  Landriau:2010cb} (see also early work in \citealt{Allen:1996wi}). In order to extend the applicability of these maps
to the small scales probed by Planck, we have used the maps described in 
\citet{Ringeval:2012tk} that have an angular resolution of $0.85'$
($\Nside=4096$). This map is obtained by considering the ISW
contribution from (\ref{eq:dtotstrg}), sourced by the Nambu-Goto
stress tensor, and which can be recast into the form \citep{Stebbins:1995}
\begin{equation}
\label{eq:isw}
\DTT(\unitn) = - {4G\mu\over c^2} \int_{\vect{X}\,\cap\,\vect{x}_\gamma} 
\left[\vect{\Xd} - 
  \dfrac{(\unitn \cdot \vect{\Xp}) \cdot \vect{\Xp}}{1 +
  \unitn \cdot \vect{\Xd}}\right] \cdot \dfrac{X \unitn - \vect{X}}{\left(X \unitn -
    \vect{X} \right)^2} \,\ud l\,.
\end{equation}
The integral is performed over all string position vectors
$\vect{X}=\{X^i\}$ intercepting our past line cone (in the transverse
temporal gauge). Primes and dots denote differentiation with respect
to the spatial and time-like worldsheet coordinates $\zeta^1$ and
$\zeta^0$ respectively, while $\ud l$ is the invariant string length
element. Taking the limit $X\unitn \rightarrow \vect{X}$ gives back
the small angle and flat sky approximation used in
\citet{Hindmarsh:1993pu, Bouchet:1988hh, Fraisse:2007nu}. For
generating the full sky map, Eq.~(\ref{eq:isw}) has been evaluated
without any other approximation and required more than $3000$
Nambu-Goto string simulations to fill the whole comoving volume
between the observer and the last scattering surface. Details on the
numerics can be found in \citet{Ringeval:2012tk}.

\begin{figure}
\begin{center}
\includegraphics[width=\hsize]{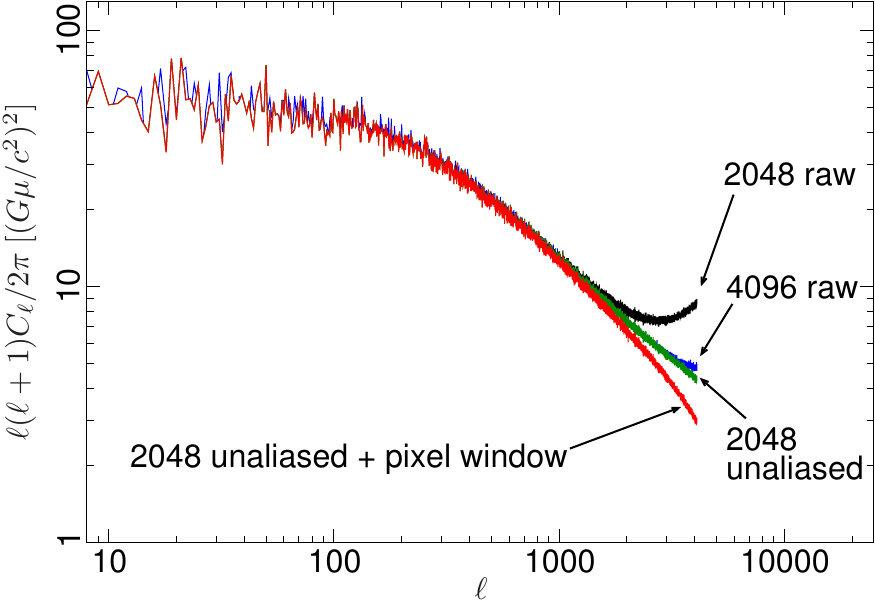}
\caption{Integrated Sachs-Wolfe angular power spectra extracted from
  the full sky cosmic string maps at different resolutions (labelled
  by $\Nside$), with or without applying the anti-aliasing procedure
  (see text). The anti-aliasing filtering gives back the correct power
  up to $\ellmax \lesssim 2 \Nside$.}
\label{fig:psrawmap}
\end{center}
\end{figure}

This method therefore includes all string effects from the last
scattering surface till today, but does not include the Doppler
contributions induced by the strings into the plasma prior to
recombination. As a result, our full sky map represents the ISW
contribution from strings, which is dominant at large and small scales
but underestimates the signal on intermediate length scales, as can
be directly checked on the power spectrum (see
Fig.~\ref{fig:psrawmap}). We therefore expect the string searches
based on this map to be less constraining than those using the power
spectrum, but certainly more robust as any line-like gravitating
object should generate such a signal.

\begin{figure*}[t]
\begin{center}
\includegraphics[width=0.925\textwidth]{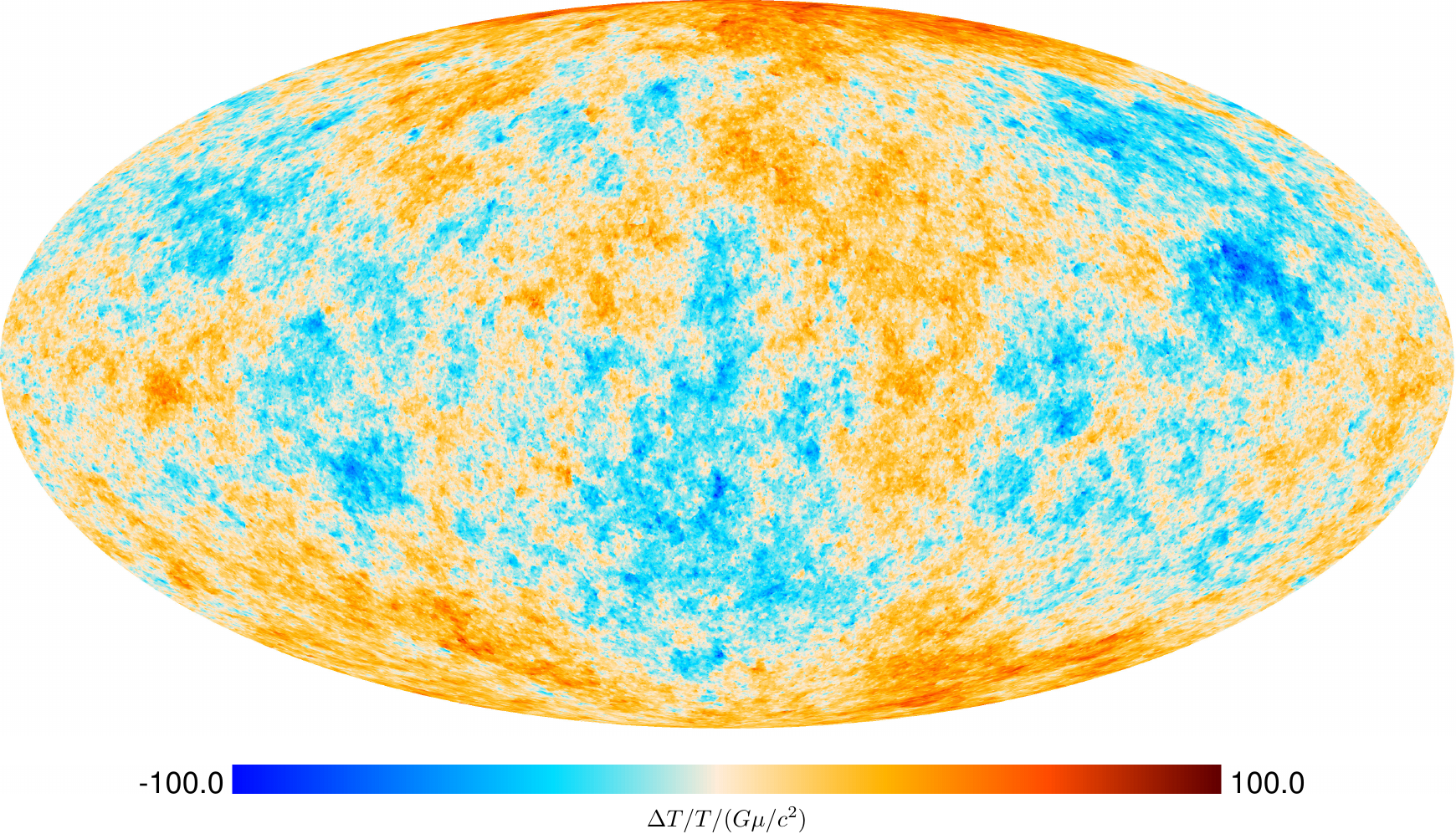}
\caption{All sky Mollweide projection of the simulated cosmic strings
  CMB sky after convolution by a Gaussian beam of $5'$ resolution. The
  color scale indicates the range of $\Delta T/T/(\Gmu)$
  fluctuations.}
\label{fig:csrawmap}
\end{center}
\end{figure*}

\begin{figure*}
\begin{center}
\includegraphics[width=0.85\columnwidth]{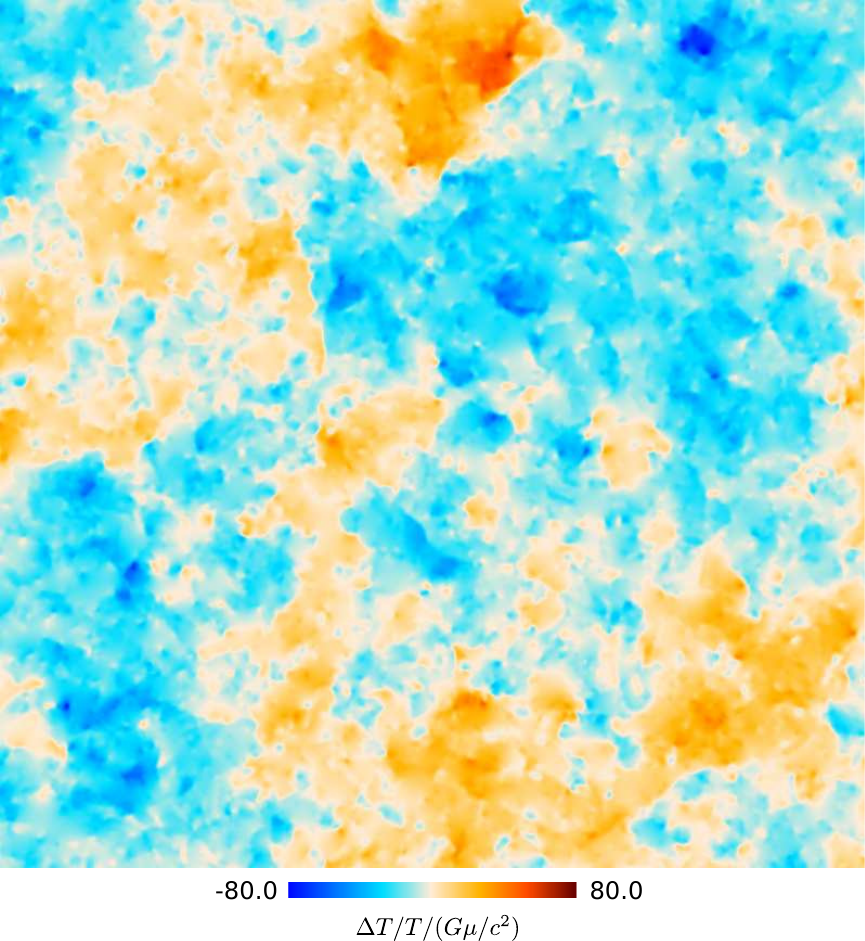}\hskip0.5in
\includegraphics[width=0.85\columnwidth]{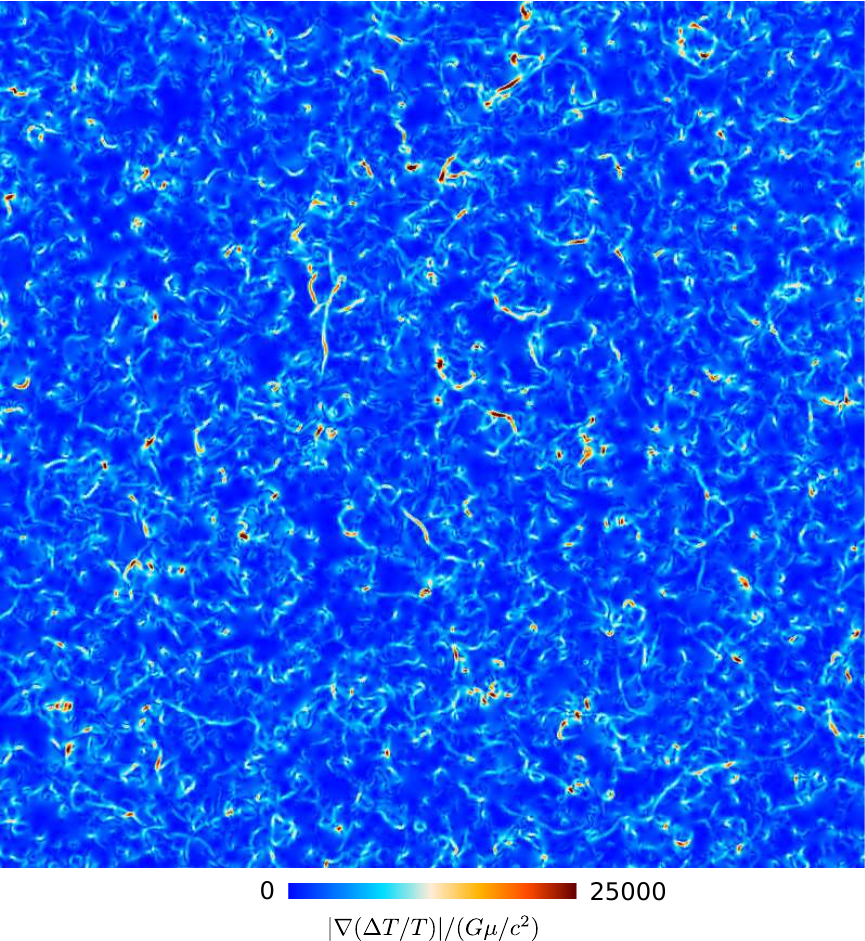}
\caption{A $20^\circ$ gnomic projection patch extracted from the full
  sky map and zooming into string induced temperature steps (see
  Fig.~\ref{fig:csrawmap}). Applying the spherical gradient magnitude
  operator enhances the temperature steps, and thus the string
  locations, even more (right).}
\label{fig:gnomrawmap}
\end{center}
\end{figure*}

Calibration and training for the non-Gaussian searches of
Sect.~\ref{sec:ngsearches} have required the generation of new full
sky and statistically independent cosmic string maps. The numerical
challenges underlying the $\Nside=4096$ map \citep{Ringeval:2012tk}
are such that it was numerically too expensive to create another one
of the same kind. At this resolution, the computations typically
require $800\,000$ cpu-hours, so we have chosen to generate three
new maps at a lower resolution of $1.7'$,
i.e., $\Nside=2048$. Unfortunately, at this lower resolution, the
simulated string maps, hereafter referred to as raw maps, exhibit a
strong aliasing at small scales that could have induced spurious
systematics even after convolution with the \Planck\ beam. This aliasing
concerns pixel-sized structures and comes from the method used to
numerically evaluate Eq.~(\ref{eq:isw}). In order to spare computing
time, the signal associated with each pixel is only computed at the
centroid direction $\unitn$. This has the effect of including some
extra power associated with string small-scale structure that is below
the pixel angular size, thereby aliasing the map. In order to address this
problem, we have used semi-analytical methods to design an optimal
anti-aliasing filter, both in harmonic space and in real space. As
discussed in \citet{Fraisse:2007nu, Bevis:2010gj}, the small scale
angular temperature power spectrum slowly decays as a power law
$\ell^{-p}$ such that any deviations from this behaviour can only
come from the aliasing. For each $\Nside=2048$ raw map, we have
performed a multi-parameter fit of the power spectrum, and of the
one-point distribution function, to extract, and then removes, its
small scale aliasing contribution. In order to validate the procedure,
we have checked that the power spectrum of each of the filtered maps
matches the one associated of the raw $\Nside=4096$ map, the latter
being also being affected but at half the scale. In
Fig.~\ref{fig:psrawmap}, we have plotted the power spectra of one of
the $\Nside=2048$ maps before and after convolution with our
anti-aliasing filter. As expected, it matches with the one extracted
from the $\Nside=4096$ map (here truncated at $\ell=4096$). Finally,
in order to include the effects associated with the $\healpix$
pixelization scheme, the anti-aliased maps have been convolved with
the $\healpix$ pixel window function before being used for further
processing.

In total, this method has provided four theoretical full sky string
maps that have been used in the string searches we will discuss in Sect.~\ref{sec:NG}. As an
illustration, we have represented in Fig.~\ref{fig:csrawmap}, one of
the filtered string map after convolution by a Gaussian beam of
$\fwhm=5'$. The color scale traces the relative temperature
anisotropies $\Delta T/T$, divided by the string tension $\Gmu$. The anisotropy patterns may look Gaussian at first
because most of the string signatures show up on the smallest length
scales. In Fig.~\ref{fig:gnomrawmap}, we have plotted a gnomic
projection representing a field of view of $20^\circ$, in which the
temperature steps are now clearly apparent. The right panel of
Fig.~\ref{fig:gnomrawmap} represents the magnitude of the spherical
gradient, which enhances the steps.

Finally, in order to provide a much larger statistical sample beyond only four
string realisations, we have also produced a collection
of 1000 small angle patches ($7.2'$) of the CMB sky
derived in the flat sky approximation~\citep{Stebbins:1988,
  Hindmarsh:1993pu, Stebbins:1995, Bouchet:1988hh,
  Fraisse:2007nu}. Although the large-scale correlations are lost, these
maps have been shown to accurately reproduce various analytically
expected non-Gaussian string effects such as the one-point and higher
$n$-points functions by \citet{Takahashi:2008ui, Hindmarsh:2009qk,
  Hindmarsh:2009es, Regan:2009hv, Yamauchi:2010ms, Yamauchi:2010vy,
  Ringeval:2010ca}.

\section{Power spectrum constraints on cosmic strings and other topological defects}

In order to compute constraints on cosmic string scenarios we just add the angular power spectrum to that for an simple adiabatic model---which assumes that they are uncorrelated--- with the fraction of the spectrum contributed by cosmic strings being $f_{10}$ at $\ell=10$. This parameter is then added as an extra parameter to the standard six parameter fit using {\tt COSMOMC} and the \Planck\ likelihood described in \cite{planck2013-p08}. We use a Flat $\Lambda$CDM cosmology defined through the physical densities of baryons, $\Omega_{\mathrm b}h^2$, and cold dark matter, $\Omega_{\mathrm c}h^2$, the acoustic scale, $\theta_{\mathrm MC}$, the amplitude, $A_{\rm s}$ and spectral index, $n_{\rm s}$ of density fluctuations and the optical depth to reionization $\tau$.  The Hubble constant is a derived parameter and is given by $H_0=100\,h\,{\rm km}\,{\rm sec}^{-1}\,{\rm Mpc}^{-1}$. We use the same  priors on the cosmological and nuisance parameters as are used in \cite{planck2013-p11} and use \textit{WMAP} polarization data to help fix $\tau$. In addition to just using the \Planck\ data, we have also added high-$\ell$ CMB data from SPT and ACT to obtain stronger constraint\citep{2013arXiv1301.0824S,2012arXiv1212.6267H}.

\begin{figure*}[!t]
\centering
\includegraphics[width=88mm]{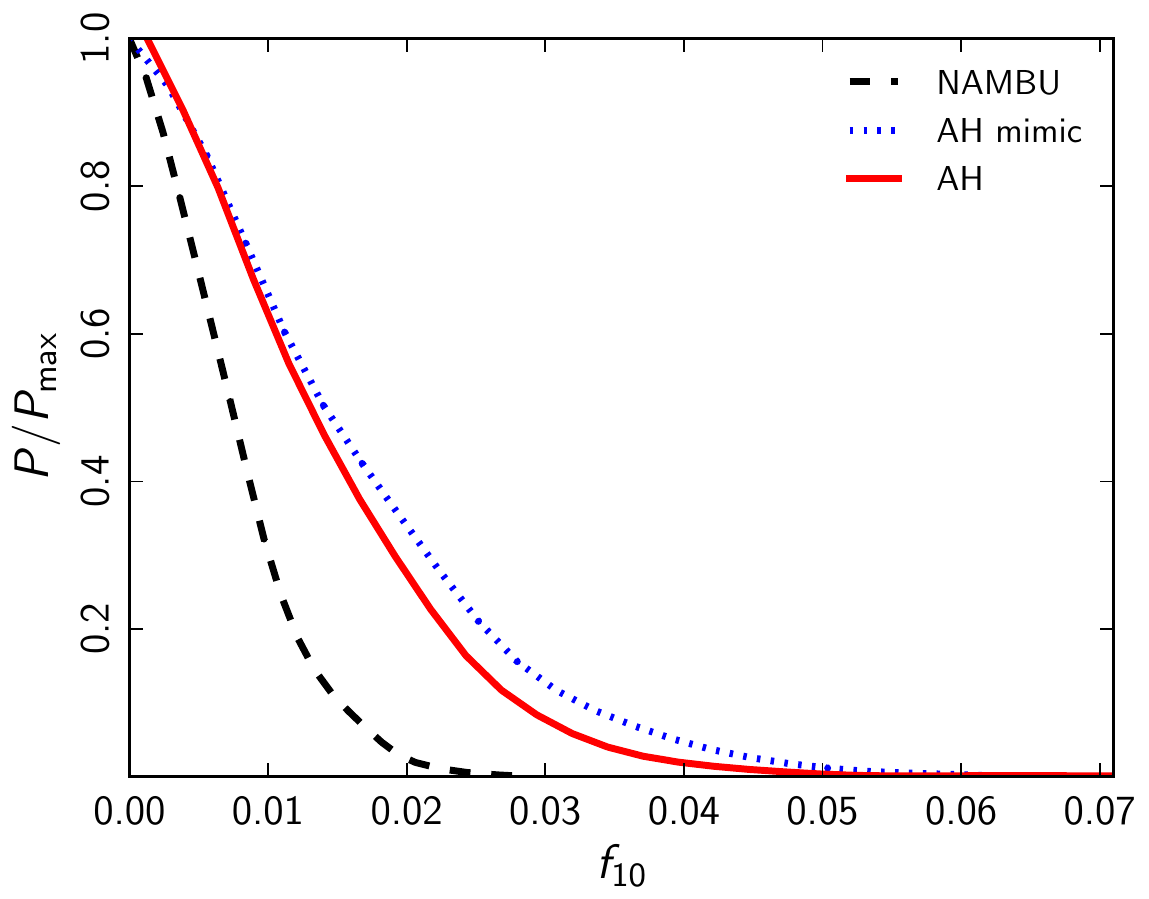}
\includegraphics[width=88mm]{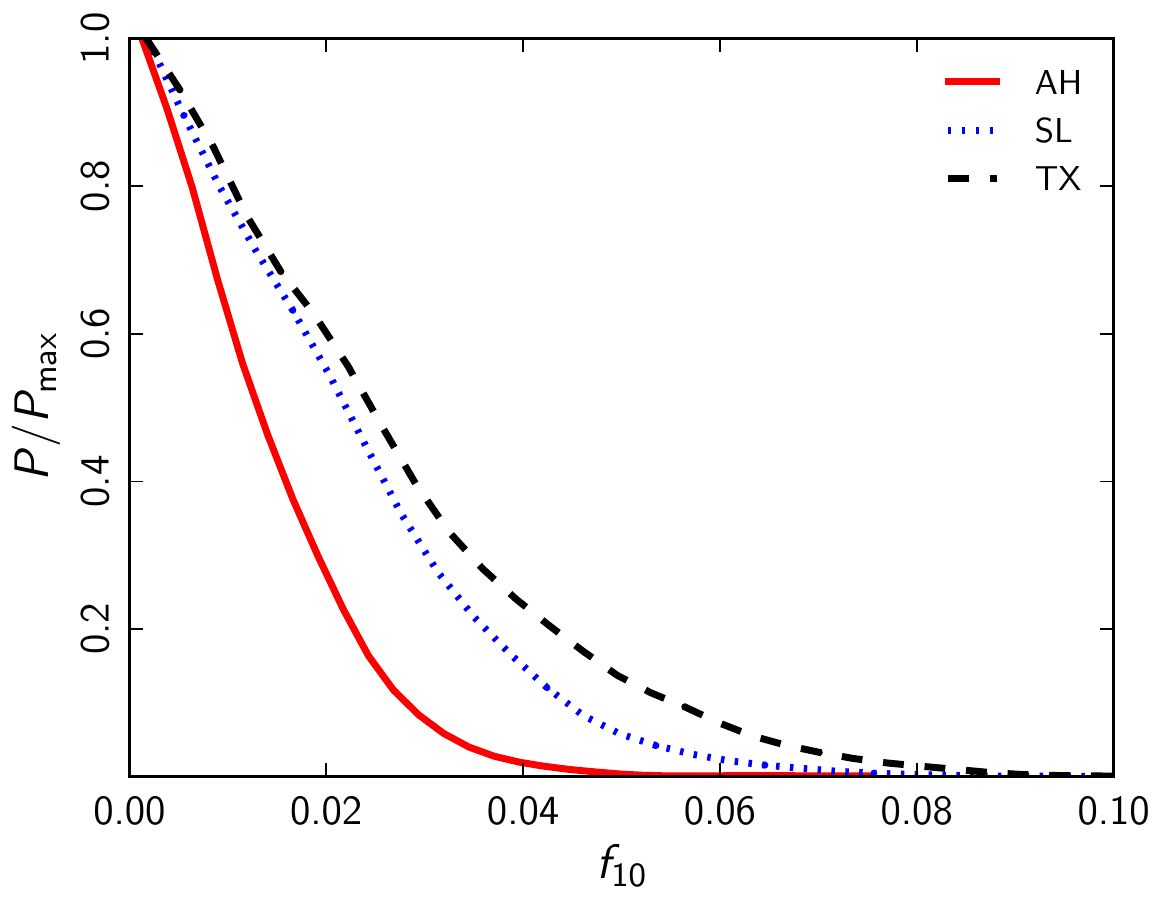} 
\caption{Marginalized constraints on $f_{10}$ for topological defects from \Planck\ data plus polarization from \textit{WMAP} (\Planck+WP). The left panel show constraints on cosmic strings, with NAMBU in black dashed, AH-mimic in blue dotted and AH in red solid. The right panel show the constraints on SL (blue dotted) and TX (black dashed) compared to AH (again solid red).}
\label{fig:f10}
\bigskip
\centering
\includegraphics[width=88mm]{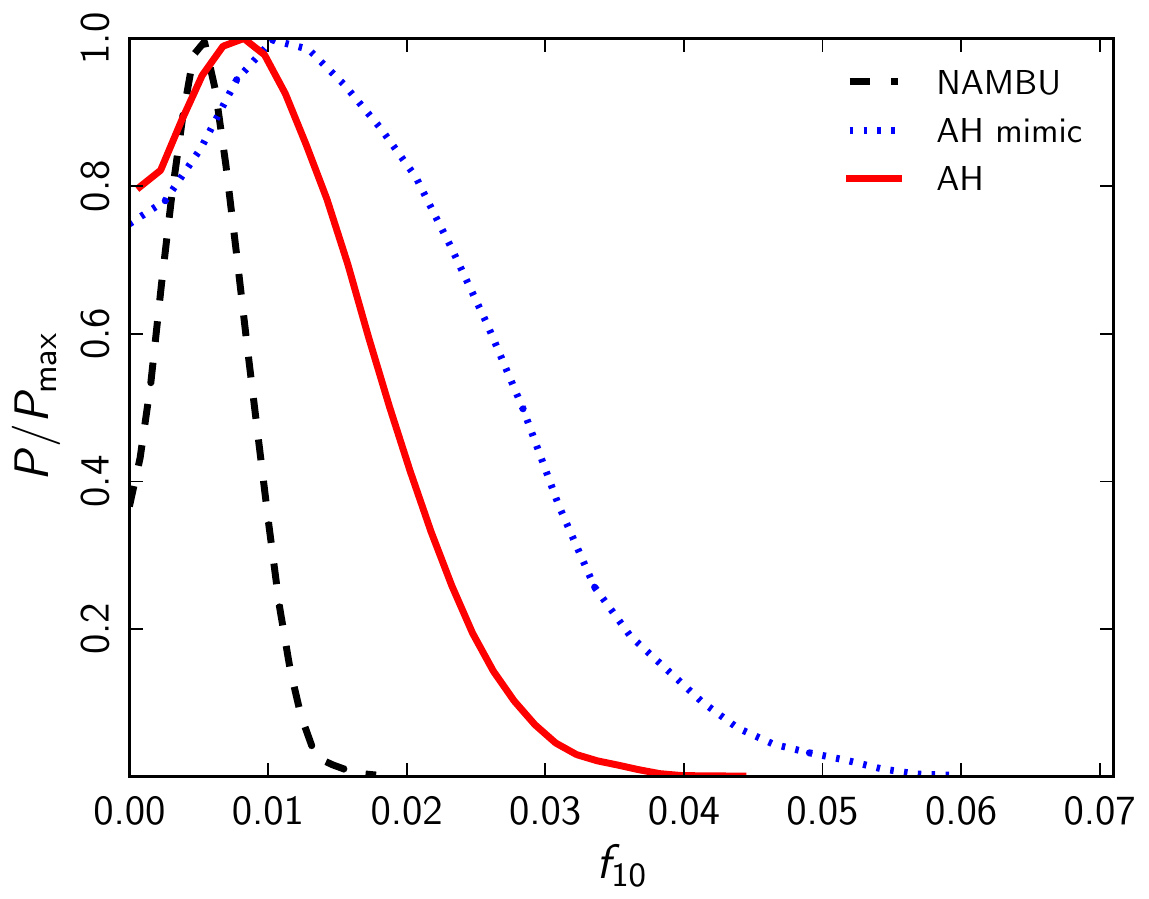}
\includegraphics[width=88mm]{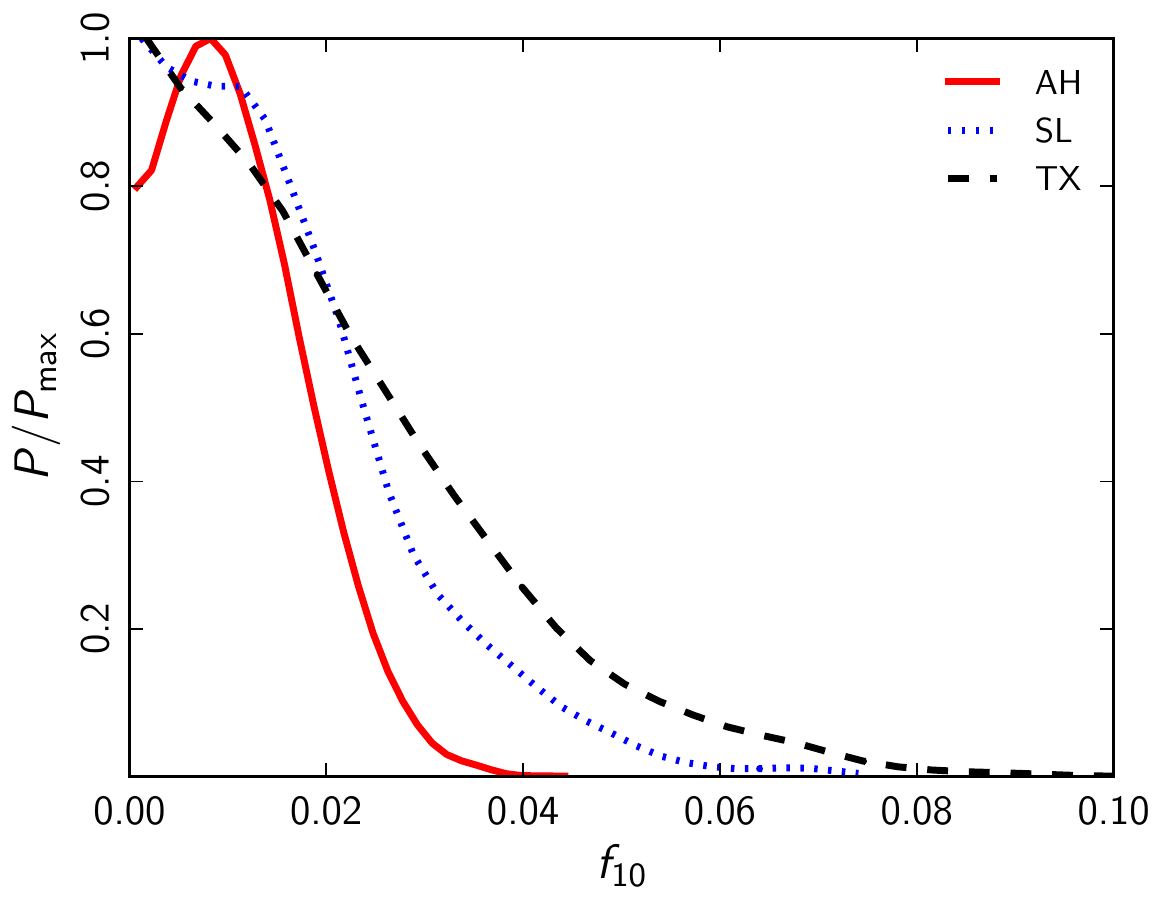}
\caption{Marginalized constraints on $f_{10}$ for topological defects with high-$\ell$ CMB data from SPT and ACT added to the \Planck\ + WP constraints data (compare with constraints shown in Fig.~\ref{fig:f10}). The left panel show constraints on cosmic strings, with NAMBU in black dashed, AH-mimic in blue dotted and AH in red solid. The right panel show the constraints on SL (blue dotted) and TX (black dashed) compared to AH (solid red).}
\label{fig:f10_highl}
\end{figure*}

For the USM-based models we use the approach used in \citet{Battye:2006pk} and \citet{Battye:2010xz}. We find that the constraints on the standard six parameters are not significantly affected by the inclusion of the extra string parameter and that there are no significant correlations with other parameters (see Table~\ref{tab:parameters}). For the case of \Planck\ data only and using the NAMBU model we find that  $\Gmu < 1.5\times 10^{-7}$ and $f_{10} < 0.015$, whereas for the AH-mimic model we find that  $\Gmu < 3.6\times 10^{-7}$ and $f_{10} < 0.033$, with all the upper limits being at 95\% confidence level. The 1D marginalized likelihoods for $f_{10}$ are presented in the upper panels of  Fig.~\ref{fig:f10}. The differences between the upper limits for the NAMBU and AH-mimic models is compatible with those seen previously using \textit{WMAP} 7-year and SDSS data \citep{Battye:2010xz}. The upper limits from this version of the \Planck\ likelihood are better than those computed from \textit{WMAP7}+SPT \citep{Dvorkin:2011aj} and \textit{WMAP7}+ACT \citep{Dunkley:2010ge} and are significantly better than those from \textit{WMAP7}+SDSS \citep{Battye:2010xz}. Based on the \Planck\ ``Blue Book'' values for noise levels we predicted \citep{Battye:2007si} a limit of $\Gmu<6\times 10^{-8}$, while the present limit is about a factor of two worse than this. The main reason for this is that the projected limit ignored the need for nuisance parameters to model high $\ell$ foregrounds and that not all the frequency channels have been used. The corresponding limits for the AH model are $f_{10}<0.028$ and $\Gmu<3.2 \times 10^{-7}$.

\begin{table}[tpb]
\begingroup
\newdimen\tblskip \tblskip=5pt
\caption{95\% upper limits on the constrained parameter $f_{10}$ and the derived parameter $\Gmu$ for the five defect models discussed in the text. We present limits using \Planck~and polarization information from \textit{WMAP} (\Planck~+ WP), and from also including high $\ell$ CMB information from ACT and SPT (\Planck~+WP+highL).}
\label{tab:limits}
\nointerlineskip
\vskip -3mm
\footnotesize
\setbox\tablebox=\vbox{
   \newdimen\digitwidth 
   \setbox0=\hbox{\rm 0} 
   \digitwidth=\wd0 
   \catcode`*=\active 
   \def*{\kern\digitwidth}
   \newdimen\signwidth 
   \setbox0=\hbox{+} 
   \signwidth=\wd0 
   \catcode`!=\active 
   \def!{\kern\signwidth}
    \halign{\hbox to 1.15in{#\leaderfil}\tabskip 1.0em&
            \hfil#\hfil&
            \hfil#\hfil&
            \hfil#\hfil&
            \hfil#\hfil\tabskip 0pt\cr
    \noalign{\doubleline\vskip 2pt}
    Defect type \hfil&\multicolumn{2}{c}{\hfil\Planck +WP\hfil} & \multicolumn{2}{c}{\hfil\Planck+WP+highL\hfill} \cr
     & $f_{10}$ & $\Gmu$ & $f_{10}$ &  $\Gmu$\cr
\noalign{\vskip 4pt\hrule\vskip 6pt}
NAMBU & 0.015 & $1.5\times 10^{-7}$ & 0.010 & $1.3\times 10^{-7}$ \cr
AH-mimic & 0.033 & $3.6\times 10^{-7}$ & 0.034 & $3.7\times 10^{-7}$\cr
\textit{AH} & 0.028 & $3.2\times 10^{-7}$ & 0.024 & $3.0\times 10^{-7}$ \cr
\textit{SL} & 0.043 & $11.0\times 10^{-7}$ & 0.041 & $10.7 \times 10^{-7}$  \cr
\textit{TX} & 0.055 & $10.6 \times 10^{-7}$ & 0.054 & $10.5 \times 10^{-7}$ \cr
 \noalign{\vskip 3pt\hrule\vskip 4pt}
}}
\endPlancktablewide

\endgroup
\end{table}

\begin{figure}[tmb]
\centering
\includegraphics[width=88mm]{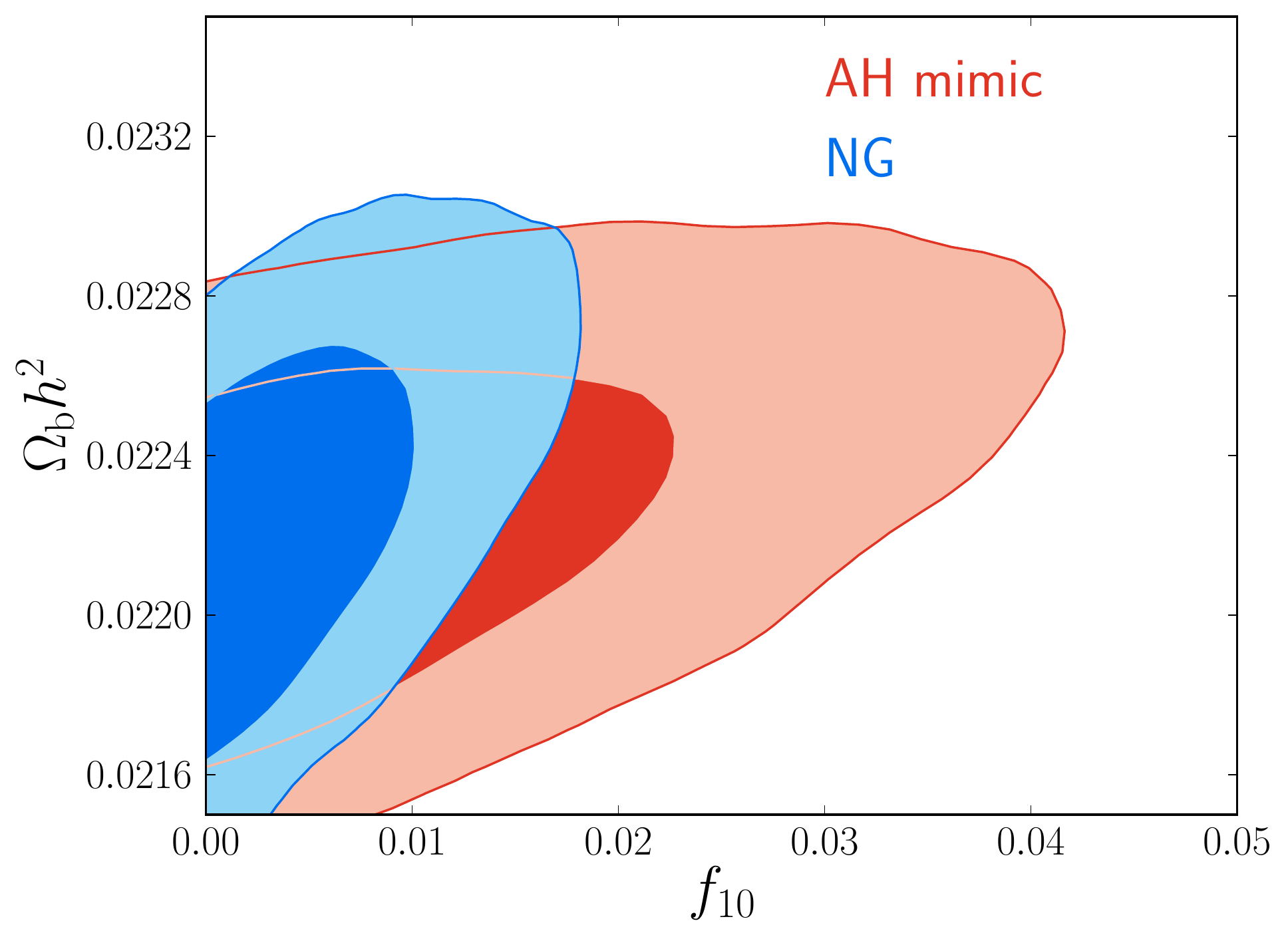}
\caption{Marginalized likelihoods for $f_{10}$  and in the $f_{10}$-$\Omega_{\mathrm b}h^2$ plane for the NAMBU model in blue and the AH \textit{mimic} model in red using \Planck\ +WP. This is the strongest correlation with any of the standard cosmological parameters }
\label{fig:marginalized}
\end{figure}

There is now very little degeneracy between the $f_{10}$ and $n_{\mathrm S}$ parameters, something that was not the case for \textit{WMAP} alone \citep{Battye:2006pk,Bevis:2007gh,Urrestilla:2011gr}. This has implication for supersymmetric hybrid inflation models as discussed in \citet{Battye:2010hg} that typically require $n_{\mathrm S}>0.98$. The simplest versions of  these models appear to be ruled out. The strongest correlation using the NAMBU and AH \textit{mimic} models is between $f_{10}$ and $\Omega_{\mathrm b}h^2$ as illustrated in Fig.~\ref{fig:marginalized}. In addition, we find agreement with \cite{Lizarraga:2012mq}, that there are significant correlations between the amount of strings $f_{10}$ in the AH model and the number of relativistic degrees of freedom $N_{\rm eff}$ as well as between $f_{10}$ and the primordial helium abundance $Y_{\rm He}$. We leave a detailed study of these correlations to later work.

\begin{table*}[!t]
\begingroup
\newdimen\tblskip \tblskip=5pt
\caption{Constraints on the fitted cosmological parameters in the case of \Planck~alone. It is clear from this that the fitted parameters are not significantly affected by the inclusion of defects.}
\label{tab:parameters}
\nointerlineskip
\vskip -3mm
\footnotesize
\setbox\tablebox=\vbox{
   \newdimen\digitwidth 
   \setbox0=\hbox{\rm 0} 
   \digitwidth=\wd0 
   \catcode`*=\active 
   \def*{\kern\digitwidth}
   \newdimen\signwidth 
   \setbox0=\hbox{+} 
   \signwidth=\wd0 
   \catcode`!=\active 
   \def!{\kern\signwidth}
    \halign{\hbox to 1.15in{#\leaderfil}\tabskip 1.0em&
            \hfil#\hfil&
            \hfil#\hfil&
            \hfil#\hfil&
            \hfil#\hfil&
            \hfil#\hfil&
            \hfil#\hfil\tabskip 0pt\cr
    \noalign{\doubleline\vskip 2pt}
 Parameter   &  NAMBU 		&  AH \textit{mimic}    & AH  & SL & TX\cr
 \noalign{\vskip 4pt\hrule\vskip 6pt}
$\Omega_{\mathrm b} h^2$  & $0.0223 \pm 0.0003$  	& $0.0223\pm 0.0003$    & $0.0223 \pm 0.0003$ & $0.0223\pm0.0003$ & $0.0223\pm0.0003$ \cr 
$\Omega_{\mathrm c} h^2$  & $0.119 \pm 0.003$ 	& $0.119 \pm 0.003$  & $0.119\pm 0.003$& $0.119\pm0.003$ & $0.119\pm0.003$ \cr
$\theta_{\mathrm MC}$  & $1.0415 \pm 0.0006$  	& $1.0415 \pm 0.0006$& $1.0415 \pm 0.0006$ & $1.0415 \pm 0.0006$ & $1.0415 \pm 0.0006$ \cr 
$\tau$ & $0.089\pm 0.013$ 		& $0.090 \pm 0.013$ & $0.090\pm0.013$ & $0.090\pm0.013$ & $0.088\pm0.014$ \cr   
$\log (10^{10} A_{\mathrm s})$   &  $3.080\pm 0.027$ & $3.080 \pm 0.026$  & $3.081\pm0.025$ & $3.081\pm0.025$ & $3.078\pm0.028$	\cr    
$n_{\rm s} $  & $0.961 \pm 0.007 $     	& $0.963 \pm 0.008$  & $0.963\pm0.008$ & $0.964\pm0.007$ & $0.965\pm0.008$  \cr
$H_0  $  & $68.4 \pm 1.3~$     	& $68.3 \pm 1.2~$  & $68.3\pm1.3~$ & $68.2\pm1.2~$ & $68.3\pm1.2~$  \cr
$G \mu /c^2$  & $ <1.5\times 10^{-7}$ 	& $ <3.6\times 10^{-7}$ & $ <3.2\times 10^{-7}$ & $<1.10 \times 10^{-6}$ & $<1.06\times10^{-6}$   \cr  
$f_{\rm 10	  }$    &	$<0.015$ 			&	$<0.033$  & $<0.028$ & $<0.043$ & $<0.055$ \cr	 
\noalign{\vskip 3pt\hrule\vskip 4pt}
}}
\endPlancktablewide

\endgroup
\end{table*}

In Fig.~\ref{fig:f10} we also present the 1D marginalized likelihoods for the texture and semilocal string models (compared to the AH field theory strings). The resulting constraints on the $f_{10}$ parameter are given in Table \ref{tab:limits} as well. For the conversion into constraints on $\Gmu$ we have that for semilocal strings $G\mu_{10}/c^2 = 5.3 \times 10^{-6}$ and for global texture $G\mu_{10}/c^2 = 4.5 \times 10^{-6}$, cf \cite{Urrestilla:2007sf}. We notice that, as expected for a fixed $G\mu$, semilocal strings lead to significantly less  anisotropies than cosmic strings (a factor of about 8 in the $C_\ell$), and texture are similar to the semilocal strings. We thus expect significantly weaker constraints on $G\mu$ for the SL and TX models, especially since in addition the constraints on $f_{10}$ for these models are weaker. Indeed we find a 95\% limit of $G \mu/c^2 < 1.10 \times 10^{-6}$ for semilocal strings and $G \mu/c^2 < 1.06 \times 10^{-6}$ for global textures.

\section{Non-Gaussian searches for cosmic strings}
\label{sec:NG}

\label{sec:ngsearches}

Cosmic strings and other topological defects generically create
non-Gaussian signatures in the cosmic microwave sky, counterparts of
their inevitable impact on the CMB power spectrum.  This is a critical
test of differentiating defects from simple inflation, while offering
the prospect of direct detection.  Searches for these non-Gaussian
defect signatures are important for two key reasons: on the one
hand, constraints from the CMB power spectrum can be susceptible to
degeneracies with cosmological parameters in the standard concordance
model; on the other hand, any apparent defect detection in the power
spectrum should have a well-defined prediction in higher-order
correlators or other non-Gaussian signals, and vice versa.
Non-Gaussian tests can also be used to distinguish cosmic defects from
residual foregrounds or systematic contributions.   Below we will present 
results from NG tests that seek strings in multipole space (bispectrum) and 
in real space (Minkowski functionals), as well as hybrid methods (wavelets).

\subsection{Foregrounds, systematics and validation}
\label{sec:challenges}

It is well-known that the microwave sky contains not only the CMB
signal but also emission from different astrophysical contaminants. In
particular, point source emission is expected to be a special cause of confusion for cosmic defects, 
notably those with high resolution signatures,
such as cosmic strings. In addition, systematic effects may also be
present in the maps at a certain level. Therefore, before claiming a
cosmological origin of a given detection, alternative extrinsic sources
 should be investigated and discarded.  This can be done by performing a number
of consistency checks in the data, most of which are common to the other non-Gaussianity papers,
where they are discussed in greater detail.   Here, we provide a brief summary of the main issues.

Foreground-cleaned CMB maps are provided
using four different component separation techniques (for further details, see
\citealt{planck2013-p06}):   \texttt{SMICA} (semi-blind approach);  
\texttt{NILC} (internal linear combination
in needlet space);\texttt{SEVEM} (internal
template fitting);  and Commander/Ruler (\texttt{C-R}, parametric method). To demonstrate the 
robustness of a particular result, it should be replicated with at least two different cleaned CMB
maps. The adoption of different masks that exclude different regions of the sky (ranging from more aggressive to
more conservative) has been used to test the stability of non-Gaussian estimates. 
Another interesting test is the use of cleaned maps at different
frequencies (for instance, those provided by the \texttt{SEVEM}
foreground separation technique). A given detection should be
consistent at all frequencies, since the behaviour of contaminants and
systematic effects will, in general, vary with frequency.
A further test is the study of noise maps constructed from the
difference between two \Planck\ maps (either at the same or at different
frequencies) smoothed to the same resolution. These maps will not
contain the CMB signal and, therefore, any NG detection should vanish on
them.  The opposite would indicate that the claimed result is due
to foreground residuals or to the presence of systematic effects.

In order to validate the various non-Gaussian methods that are 
described below, we instituted a series of \textit{Planck String
  Challenges}.  These were blind tests employing post-recombination
string simulation maps with an unknown $\Gmu$, co-added to a Gaussian
CMB map, together with standard masks and increasingly more realistic
noise models.  For calibrating the non-Gaussian tests, several
different string simulations were also provided.  In addition, 1000
$\Lambda$CDM Gaussian maps, with simulated noise and using the same mask, 
were provided for analysis purposes, notably for determining the variance of 
different techniques.  The aim has been to determine the sensitivity of the
proposed non-Gaussian tests and to see if the $\Gmu$ in the string
challenge map could be measured accurately.  The results from these
challenges were an important part of the validation for each of the methods 
described below.  \Planck\ simulation pipelines for each of the component
separation methods were also used to estimate realistic foreground residuals 
and were used to validate non-Gaussian pipelines and to remove string 
signal bias.  
 
\subsection{Cosmic string bispectrum}

\def\lall{\ell_1,\ell_2,\ell_3}
\def\blll{b_{\ell_1\ell_2\ell_3}}
\def\bslll{b^{\rm string}_{\ell_1\ell_2\ell_3}}
\def\bPSlll{b^{\rm PS}_{\ell_1\ell_2\ell_3}}
\def\bISWlll{b^{\rm ISW}_{\ell_1\ell_2\ell_3}}
\def\hlll{h_{\ell_1\ell_2\ell_3}}
\def\lmax{\ell_\textrm{max}}
\def\barQ{{Q}}
\def\barQn{\barQ_n}
\def\barQm{\barQ_m}
\def\barQp{\barQ_p}
\def\baQn{\baQ_n}
\def\baQp{\baQ_p}
\def\baQ{{\alpha}^{\scriptscriptstyle{Q}}}
\def\bbQ{{\beta}^{\scriptscriptstyle{Q}}}
\def\bbQn{\bbQ_n}
\def\barR{R}
\def\barRn{\barR_n}
\def\barRm{\barR_m}
\def\barRp{\barR_p}
\def\aR{\alpha^{R}}
\def\aRn{\aR_n}
\def\aRp{\aR_p}
\def\baR{{\alpha}^{\scriptscriptstyle{R}}}
\def\baRn{\baR_n}
\def\baRp{\baR_p}
\def\bbR{{\beta}^{\scriptscriptstyle{R}}}
\def\bbRn{\bbR_n}
\def\un{{\bf \hat{n}}}
\def\fnl{f_{\rm NL}}
\def\Fnl{F_{\rm NL}}

\subsubsection{Modal bispectrum methods}

The CMB bispectrum is the three point correlator of the $a_{lm}$ coefficients, $B^{\ell_1 \ell_2 \ell_3}_{m_1 m_2 m_3} = a_{\ell_1 m_1} a_{\ell_2 m_2} a_{\ell_3 m_3}$.   For the purposes of a search for cosmic strings we assume the network cumulatively creates a statistically isotropic signal, that is, we can employ the angle-averaged reduced bispectrum $\blll$, defined by
\eq
b_{\ell_1 \ell_2 \ell_3} &=& \sum_{m_i}\hlll^{-2}  {\cal G}^{\ell_1 \ell_2 \ell_3}_{m_1 m_2 m_3} B^{\ell_1 \ell_2 \ell_3}_{m_1 m_2 m_3}\,,
\qe
where $\hlll$ is a weakly scale-dependent geometrical factor and $ {\cal G}^{\,\,l_1\; l_2\; l_3}_{m_1 m_2 m_3}$ is the well-known Gaunt integral over three $Y_{\ell m}$s that can be expressed in terms of Wigner-$3j$ symbols. 
The CMB bispectrum $\blll$ is defined on a tetrahedral domain of 
multipole triples $\{\ell_1 \ell_2 \ell_3\}$ satisfying both a triangle condition and 
$\ell<\lmax$ set by the experimental resolution.   When seeking the string bispectrum $\bslll$ in the \Planck\ data, we employ the following estimator to find or limit its amplitude:
\eq\label{eq:approxestimator}
{\cal E} = \frac{1}{\tilde{N}^2} \sum_{l_i m_i} \frac{{\cal G}^{\ell_1 \ell_2 \ell_3}_{m_1 m_2 m_3} \, \bslll }{ \tilde{C}_{\ell_1}\tilde{C}_{\ell_2}\tilde{C}_{\ell_3} } \,a_{\ell_1 m_1} a_{\ell_2 m_2} a_{\ell_3 m_3}\,,
\qe
where we assume a nearly diagonal covariance matrix $C_{\ell_1 m_1, \ell_2 m_2} \approx C_\ell \,\delta_{\ell_1\ell_2}\,\delta_{m_1\,-m_2}$ and we modify $C_\ell$ and $\blll$ appropriately to incorporate  instrument beam and noise effects, as well as a cut-sky.  To simplify \eqref{eq:approxestimator},  we have ignored the ``linear term'' (which is included in the analysis).    A much more extensive introduction to bispectrum estimation can be found in \cite{planck2013-p09a}.   

A key step in observational searches for non-separable bispectra, such as those induced by cosmic strings (denoted by $\bslll$), is to expand it into separable modes \citep{Fergusson:2008ra,Fergusson:2009nv} taking the signal-to-noise-weighted form:
\eq \label{eq:cmbestmodes}
\frac{\bslll }{\sqrt{C_{\ell_1}C_{\ell_2}C_{\ell_3}}} \, = \sum_n \beta^Q_n\, \barQn(\lall)\,,
\qe 
where the modes $\barQn (\lall)= {\textstyle \frac{1}{6}}[\bar q_p(l_1)\, \bar q_r(l_2)\, \bar q_s(l_3) +  \hbox{perms.}]$ are constructed from symmetrized products (the $n$ label a distance-ordering for the triples $\{prs\}$).     The product basis functions $\barQn (\lall) $  are not in general orthogonal,
so it is very useful to construct a related set of orthonormal mode functions $\barRn(\lall)$  such that 
$\langle  \barRn,\,\barRp\rangle = \delta_{np}$.   
Substituting the separable mode expansion (\ref{eq:cmbestmodes}) reduces the estimator (\ref{eq:approxestimator}) to the simple form
\eq\label{eq:estimatorsum}
{\cal E} = \frac{1}{N^2} \sum_n \beta^Q_n\, \beta^Q_n\,, 
\qe
where the  $\bbQn$ coefficients are found by integrating products of three \Planck\ maps filtered using the basis function $\baQn$ (an efficient product with each map multiplied by the separable $q_r(\ell)$).   
We can validate this estimator by using the modal methodology to create CMB map realisations for cosmic strings from the predicted $\beta_n^Q$ with a given $\Gmu$ (see \citealt{Fergusson:2009nv}).
It is easy to show 
that the expectation value for $\beta_n^R$ for such an ensemble of maps in the orthogonal basis should be 
\eq\label{eq:bestfitbeta}
\langle \beta^R_n\rangle =  \alpha^R_n\,.
\qe 
Alternatively, we can exploit this fact by reconstructing the $\baRn$ from given CMB map realisations created directly from string simulations, an approach we will adopt here.

\subsubsection{Post-recombination string bispectrum}

\begin{figure}
\begin{center}
\includegraphics[width=8.8cm]{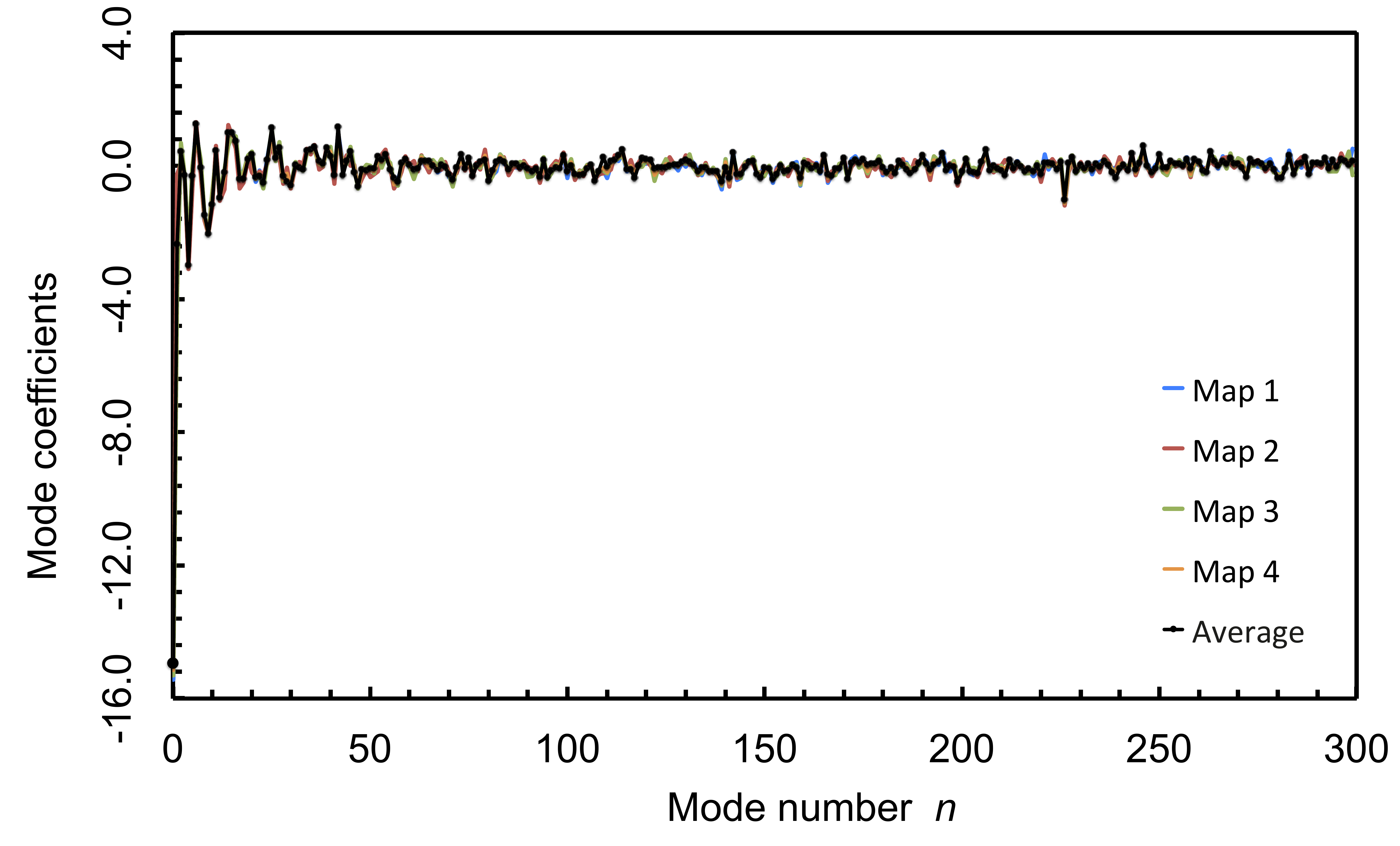}
\end{center}
\begin{center}
\includegraphics[width=8.8cm]{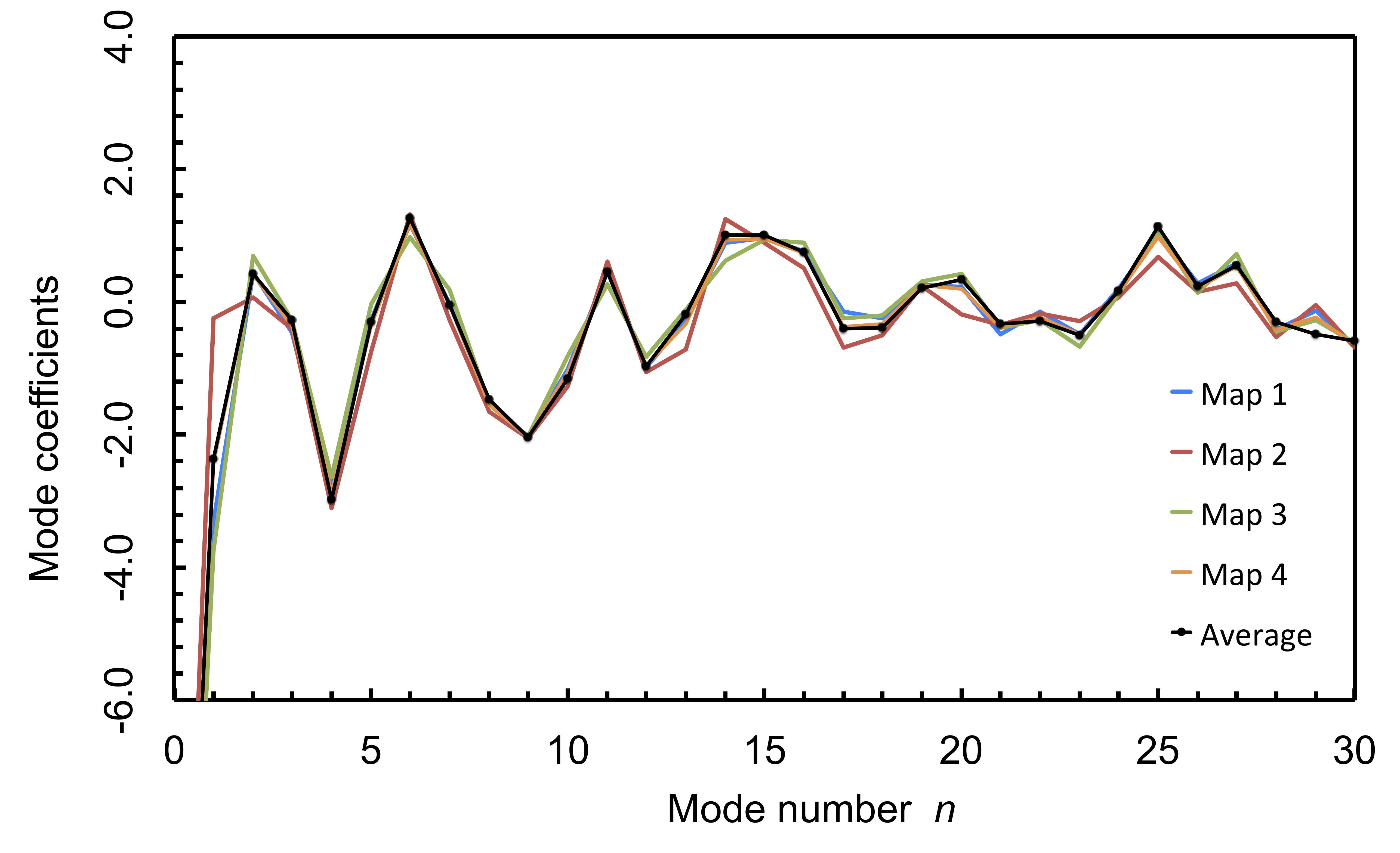}
\end{center}
\caption{Coefficients $ \baRn$  \eqref{eq:cmbestmodes}  for the hybrid Fourier mode expansion of the cosmic string bispectrum  \eqref{eq:cmbestmodes}.   The average value $\langle \baRn\rangle$ (bold black line) is in remarkable agreement with all four string simulations as can be seen for $n<30$ (lower panel), with each exhibiting better than a 97\% correlation overall. 
{\it Figure to be updated imminently.} ]
}
\label{fig:stringbispectrummodes}
\end{figure}

In order to estimate the string bispectrum at \Planck\ resolution we employed the modal reconstruction method \eqref{eq:bestfitbeta} on the post-recombination string challenge simulations described in Sect~2.5.  These string maps include the accumulated line-like discontinuities induced by the string network on CMB photons propagating from the surface of last scattering to the present day.   This work does not include recombination physics, that is, contributions from the surface of last scattering that will increase the string anisotropy signal substantially.   As discussed previously, there are four full-sky maps of two different resolutions, which were provided for the purpose of calibrating \Planck\ searches for cosmic strings \citep{Ringeval:2012tk}.   For the modal analysis, we have adopted the hybrid polynomial basis augmented with a local shape mode (in total with $n_{\rm max}=600$ modes), as well as a hybrid Fourier basis ($n_{\rm max}=300$),  which are both described in \cite{planck2013-p09a}. 

To extract the string bispectrum in a \Planck-realistic context, we chose a fairly high non-Gaussian signal with $\Gmu = 1\times 10^{-6}$.  The normalized string maps were added to noise maps generated by the component separation pipelines of \texttt{SMICA}, \texttt{NILC} and \texttt{SEVEM},  creating twelve sets of 200 simulated string realisations.  Each of these maps was then filtered using the modal estimator to find the $\beta^R_n$ coefficients appropriate for each component-separation method.   After averaging each set of modal coefficients  $\alpha^R_n= \langle \beta^R_n\rangle$
 over the different (unmasked) noise realisations, we found remarkable consistency between the estimated $\beta^R_n$ for the four string simulations as shown in Fig.~\ref{fig:stringbispectrummodes} for the Fourier modes.   Agreement was good across all the 300 $\alpha^R_n$ modes determined, as shown in detail for $n$$=$$1$$-$$30$ (see the lower panel of Fig.~\ref{fig:stringbispectrummodes}).  

Quantitatively, the different string simulations produced bispectrum shapes that had above 97\% correlations with each other (i.e., $\sum_nN^{-1}\alpha^{(1)}_n \,\alpha^{(2)}_n > 0.97$ for $N^2=\sum_n\alpha^{(1)\,2}_n \,\sum_n \alpha^{(2)\,2}_n $). The overall integrated bispectrum amplitudes was  consistent to within 4\%.  Despite only four string map simulations, these are small errors relative to experimental and theoretical uncertainties.  This robustness indicates that the overall string bispectrum signal at \Planck\ resolution is a statistical summation of very many contributions from the millions of strings between the observer and the last scattering surface.   To ensure the bispectrum  $C_l$ weighting was not significantly affected by the presence of the large string signal, we repeated the modal extraction procedure for $\Gmu = 5\times 10^{-7}$ (the string bispectrum amplitude reduced by a factor of 8).  For the same string simulation, the shape correlations for different $\Gmu$ were 99.4\% or above and the amplitude scaled as expected with  $(\Gmu)^3$ to within 2\%.  The string bispectrum shown in Fig.~\ref{fig:stringbispectrummodes} is well converged with random errors from the averaging procedure small relative to the actual signal.   We conclude that, assuming the physics and numerical accuracy of the string simulations that are available, we have extracted a string bispectrum of sufficient accuracy for the present non-Gaussian analysis. 

\begin{figure}
\begin{center}
\includegraphics[width=8.8cm]{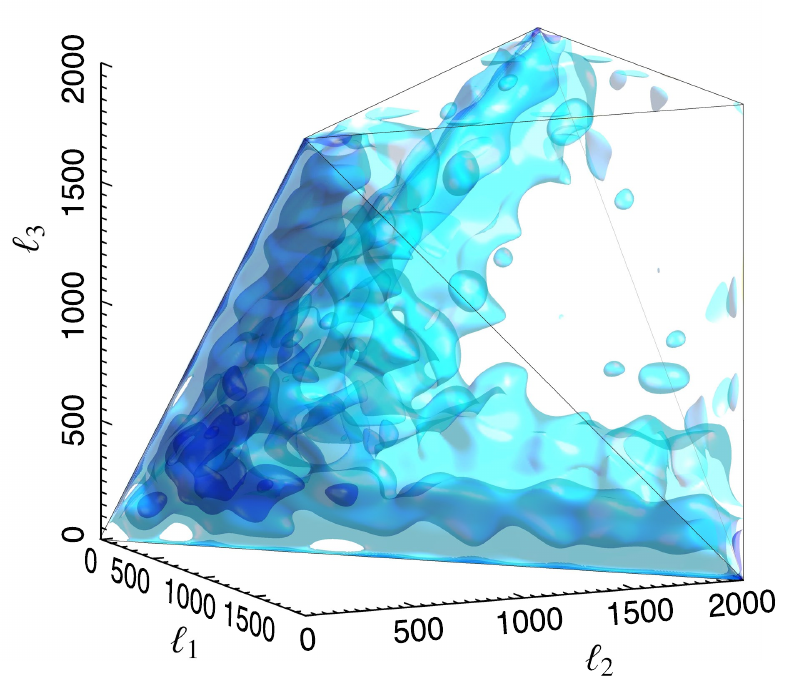}
\end{center}
\caption{Modal reconstruction of the post-recombination string bispectrum  \eqref{eq:cmbestmodes} extracted from \Planck\ resolution map simulations. This is a 3D view of the allowed tetrahedral set of multipoles $(\lall)$ showing isosurfaces of the bispectrum density with darker blue for more negative values (it is normalized relative to the constant SW bispectrum). }
\label{fig:stringbispectrum}
\end{figure}

The overall three-dimensional reconstruction of the string bispectrum shape is shown in Fig.~\ref{fig:stringbispectrum}, normalized in the usual way to approximately illustrate the signal-to-noise expected (that is, removing an overall $l^{-4}$ scaling, by dividing by the constant Sachs-Wolfe bispectrum shape).   The first point to note is that the bispectrum is negative, reflecting the underlying string velocity correlations and curvature correlations that have created it;  in the expanding universe, curved strings preferentially collapse, creating a negative temperature fluctuation towards the centre and a positive signal outside.   In the overall spectrum, the $n$$=$$0$ mode is dominant, but it is modulated by other modes providing further interesting structure that could be described as broad arms extending along each axis (see Fig.~\ref{fig:stringbispectrummodes}); although somewhat ``squeezed'',  the correlation with the local model is low.   The string simulation power spectrum shown in  Fig.~\ref{fig:psrawmap} can be understood to be quantitatively  modulating the string bispectrum away from the axes, with the signal slowly decaying beyond (say) $l_1,l_2$$\,>\,$$500$ in the $l_3$ direction.    

The CMB bispectrum and trispectrum induced by the post-recombination gravitational effects of cosmic strings have been estimated analytically \citep{Hindmarsh:2009qk, Hindmarsh:2009es,Regan:2009hv}.   With simplifying assumptions, these predicted  that the constant mode would be dominant with a broad central ``equilateral'' peak,  but not the substructure observed in Fig.~\ref{fig:stringbispectrum}.    In terms of missing physics in this post-recombination string bispectrum, we expect the recombination signal to lie in the range $\ell=200-1000$ (shown in the full NAMBU power spectrum in Fig.~\ref{fig:comp}) and to significantly enhance the the overall amplitude of the bispectrum (see also \citealt{Landriau:2010cb}, where recombination physics is included).   

The correlation of the post-recombination string bispectrum with standard primordial shapes is small, because it does not contain an oscillatory component from the transfer functions.  The local, equilateral and orthogonal bispectrum models correlate with strings at about 6\%, 11\% and 12\%, respectively.   The late-time CMB  ISW-lensing bispectrum $\bISWlll$  is also mainly negative, but it is much more squeezed/flattened and correlates at only about 11\% with the present string bispectrum, so the predicted ISW signal should provide a small positive bias for string detection of about 0.44$\,\sigma$ (see below).   Diffuse point-source contamination is of greater concern because at $\lmax =2000$ this has a 40\% anti-correlation with strings (for the simple Poisson-distributed point source template with $\bPSlll$$=$$\,\hbox{constant}$).  This close relationship with point sources requires a joint analysis.   Other foreground contamination must also be considered, as we shall discuss,  and for this we rely on realistic simulated foreground residuals provided by the \Planck\ component separation pipelines (see \citealt{planck2013-p06}).

\subsubsection{\Planck\ string bispectrum results}
\label{sec:bispectrumresults}

Using the modal bispectrum estimator, we have searched for the string bispectrum in the \Planck\ CMB maps obtained using the foreground-separation techniques \texttt{SMICA}, \texttt{NILC} and \texttt{SEVEM}.     We note that the modal estimator has passed through the full validation suite of NG tests described in the \cite{planck2013-p09a}, where further details about the analysis can be found.  In summary, we have used the standard U73 mask, which includes a Galactic cut and a conservative point source mask, together with  ``inpainting'' as in the $\fnl$ analysis (essentially apodizing the mask).  Together with the foreground separated maps, realistic noise simulations were were used to determine the estimator's linear correction term and to determine the bispectrum variance, which was very nearly optimal.   For calibration purposes, we always compare to the string model with tension $\Gmu = 1\times 10^{-6}$, for which can choose to define and normalise a  string bispectrum parameter $\fnl=1$ at $\lmax =2000$.   The standard deviation $\Delta \fnl =0.2$ obtained from this \Planck\ analysis would imply $\Gmu = 1\times 10^{-6}$ string detection at $5\sigma$.  Instead, defining an $\Fnl$ normalized relative to the local model, we could expect to measure $ \Fnl = 31.6\pm 6.3$.   The strong scaling of the bispectrum amplitude on the string tension  $\propto (\Gmu)^3$, implies a given measurement yields $\Gmu = (\fnl)^{1/3}\times 10^{-6}$.

The results of the string bispectrum estimation for each of the  \texttt{SMICA}, \texttt{NILC} and \texttt{SEVEM} maps are shown in Table~\ref{tab:bispectrum}.  Given that the \Planck\ data exhibit significant detections of both ISW lensing and residual point source signals, we also quote their measurements. In the first instance, we have undertaken an independent analysis of each map, which showed no evidence for a cosmic string signal (all estimates were were within $1\sigma$).   However, given the significant measurement of diffuse point sources and their strong anti-correlation with the string bispectrum, we have also undertaken a joint analysis of the \Planck\ data (which in this case is the same as marginalising over the point source signal).   Before doing so, we subtract the expected ISW lensing signal that provides a $\fnl = 0.09$ string bias.   The third column in Table~\ref{tab:bispectrum} shows that marginalising point sources enhances the string signal up to the $2\sigma$ level for all component separation methods; essentially the constant mode becomes more strongly negative once the measured point sources are removed.   

Other foregrounds, such as dust emission, could potentially produce spurious string signals if not subtracted properly.   Foreground contamination has been extensively simulated within the \Planck\ pipeline and foreground residual maps are provided by each component separation team.   We have analysed the residual maps provided by both \texttt{SMICA} and \texttt{NILC} to seek evidence of correlations with the string bispectrum.  The \texttt{SMICA} combined-residual map, without point sources and analysed with realistic noise, produces a string bias of $\fnl = 0.23$, which after ISW subtraction becomes  $\fnl = 0.14$ (relative to a variance $\Delta \fnl =0.20$).  After both ISW and foreground residual subtraction, a joint analysis with point sources yields a \texttt{SMICA} string signal $\fnl = 0.37 \pm 0.21$ .   The apparent string bias in \texttt{NILC} from residual foregrounds was even higher $\fnl = 0.22$ (after ISW subtraction), meaning a joint analysis obtained  $\fnl = 0.23 \pm 0.21$ (see Table~\ref{tab:bispectrum}).

\begin{table}[tmb]
\begingroup
\newdimen\tblskip \tblskip=5pt
\caption{Modal bispectrum analysis of foreground-separated \texttt{SMICA}, \texttt{NILC} and \texttt{SEVEM} maps showing $\fnl$ from strings, ISW-lensing and diffuse point sources. Three values for $\fnl$ are given from independent analysis,  joint point source/string analysis after ISW-lensing subtraction, and joint analysis after both ISW-lensing and foreground residual subtraction.   Resulting 95\% confidence limits for $\Gmu$ are also given.}
\label{tab:bispectrum}
\nointerlineskip
\vskip -3mm
\footnotesize
\setbox\tablebox=\vbox{
   \newdimen\digitwidth 
   \setbox0=\hbox{\rm 0} 
   \digitwidth=\wd0 
   \catcode`*=\active 
   \def*{\kern\digitwidth}
   \newdimen\signwidth 
   \setbox0=\hbox{+} 
   \signwidth=\wd0 
   \catcode`!=\active 
   \def!{\kern\signwidth}
    \halign{\hbox to 0.32in{#\hfil}\tabskip 1.0em& 
   \hbox to 0.7in{#\hfil}&
            \hfil#\hfil&
            \hfil#\hfil&
            \hfil#\hfil\tabskip 0pt\cr
    \noalign{\doubleline\vskip 2pt}
    \hfil&   \hfill Bispectrum \hfill  & Independent  & ISW subtract &  ISW/FG res. \cr
     Method &\hfill signal type   \hfill & analysis $\fnl$ & Joint $\fnl$ &  Joint $\fnl$\cr
\noalign{\vskip 4pt\hrule\vskip 6pt}
\texttt{SMICA} \hfil & Lensing ISW & $0.75\pm 0.37$ & --  & -- \cr
 \hfil &Diff.~PS$\,\times10^{28}$ & $1.05\pm 0.32$ & $1.35\pm 0.34$ & $1.40 \pm 0.34$ \cr
\hfil & Cosmic strings & $0.19 \pm 0.20$ & $0.50\pm 0.21$  & $0.37 \pm 0.21$   \cr
\hfil & $\Gmu$ (95\%) & $8.4 \times 10^{-7}$ & $9.7 \times 10^{-7}$ & $9.3 \times 10^{-7}$ \cr
 \noalign{\vskip 3pt\hrule\vskip 0pt}\cr
\texttt{NILC} \hfil & Lensing ISW & $0.91\pm 0.36$ & --  & -- \cr
 \hfil &Diff.~PS$\,\times10^{28}$ & $1.16\pm 0.32$ & $1.44\pm 0.34$ & $1.44\pm 0.34$\cr
\hfil & Cosmic strings & $0.13 \pm 0.20$ & $0.46\pm 0.21$  & $0.23\pm 0.21$ \cr
\hfil & $\Gmu$ (95\%) & $8.1 \times 10^{-7}$ & $9.6 \times 10^{-7}$ & $8.7 \times 10^{-7}$\cr
 \noalign{\vskip 3pt\hrule\vskip 0pt}\cr
\texttt{SEVEM} \hfil & Lensing ISW & $0.6\pm 0.36$ & --  & -- \cr
 \hfil &Diff.~PS$\,\times10^{28}$ & $1.07\pm 0.35$ & $1.33\pm 0.38$ & -- \cr
\hfil & Cosmic strings & $0.10 \pm 0.20$ & $0.38\pm 0.21$  & --  \cr
\hfil & $\Gmu$ (95\%) & $7.9 \times 10^{-7}$ & $9.3 \times 10^{-7}$ & -- \cr
 \noalign{\vskip 3pt\hrule\vskip 4pt}
}}
\endPlancktablewide

\endgroup
\end{table}

We conclude, given our present understanding of point sources and foregrounds, that there does not appear to be significant evidence for a string bispectrum signal in the \Planck\ nominal mission maps, so we infer the following post-recombination bispectrum constraint on strings (from $\fnl = 0.30 \pm 0.21$):
\eq\label{eq:bispectrumconstraint}
\Gmu < 8.8\times 10^{-7}  \qquad \hbox{(95\% confidence)}\,. 
\qe
The susceptibility of the string bispectrum to point source and other foreground contamination deserves further investigation and will require improved characterisation of the diffuse point source bispectrum (beyond the simple Poisson model), as well as identification of other foreground residuals generating a small string bias.   

The string bispectrum constraint \eqref{eq:bispectrumconstraint} is a conservative upper limit on the string tension $\Gmu$ because we have not included recombination contributions.   Although this constraint is weaker than that from the power spectrum, it is an independent test for strings and the first quantitative string bispectrum limit to date.   This should be considerably improved in future by inclusion of recombination physics and more precise foreground analysis.  A comparison with the power spectrum amplitude indicates the string bispectrum should rise by  $(2)^{3/2}$, which, together with the full mission data, would see the sensitivity improve by a factor of two (allowing constraints around $\Gmu < 4\times 10^{-7} $).   We note that the bispectrum is not the optimal non-Gaussian test for strings, because the string signal is somewhat suppressed by symmetry (the bispectrum cancels for straight strings).  This fact motivates further study of the trispectrum, for which the \Planck\ sensitivity is forecast to be 
$\Delta \Gmu  \approx 1 \times 10^{-7}$ \citep{Fergusson:2010gn}, as well as joint analysis of polyspectra.

\subsection{Steerable wavelet searches for cosmic strings}

Wavelets offer a powerful signal analysis tool due to their ability to localise signal content in scale (cf.\ frequency) and position simultaneously.  Consequently, wavelets are well-suited for detecting potential CMB temperature contributions due to cosmic strings, which exhibit spatially localised signatures with distinct frequency content. Wavelets defined on the sphere are required to analyse full-sky \Planck\ observations  (see, for example, \citealt{freeden:1997a,wiaux:2005,sanz:2006,mcewen:2006:cswt2,starck:2006,marinucci:2008,wiaux:2007:sdw}).  

We perform an analysis using the steerable wavelets on the sphere constructed by \citet{wiaux:2005}.  Here we exploit steerability to dynamically adapt the orientations analysed to the underlying data, performing frequentist hypothesis testing.  
We apply the first (1GD) and second (2GD) Gaussian derivative steerable wavelets, defined on the sphere through a stereographic projection, in order to search for cosmic strings in the {\it Planck} data. A steerable wavelet is a directional filter whose rotation by $\chi \in [0,2\pi)$ about itself can be expressed in terms of a finite linear combination of non-rotated basis filters. Thus, the analysis of a signal with a given steerable wavelet $\Psi$ naturally identifies a set of wavelet coefficients, $W_{\Psi}(\omega_0,\chi,R)$, which describe the local features of the signal at each position $\omega_0$ on the sphere, for each orientation $\chi$ and for each physical scale $R$. 
Several local morphological properties can be defined in terms of the wavelet coefficients \citep{Wiaux08}, including the signed-intensity,
\begin{equation}\label{eq:swr_1}
I\left(\omega_{0},R\right)\equiv W_\Psi\left(\omega_{0},\chi_{0},R\right)\,.
\end{equation}
This quantity represents the value of the wavelet coefficient at the local orientation $\chi_{0}\left(\omega_0,R\right)$ that maximizes the absolute value of the wavelet coefficient itself. 

The presence of a cosmic string signal in the CMB is expected to leave a non-Gaussian signature that induces a modification in the distribution of $I(\omega_0,R)$ with respect to the lensed Gaussian case. We calibrated the dependence of these signatures on the string tension using four simulations of the cosmic string contribution \citep{Ringeval:2012tk} combined with a large set of lensed Gaussian CMB realizations, along with a realistic description of the {\it Planck} 
instrumental properties  (refer to \cite{planck2013-p06}). 

A wide range of string tension values were explored, $\Gmu \in [2.0 \times 10^{-7}, 1.0 \times 10^{-6}]$, considering several wavelet scales, $R=\left[4.0,4.5,5.0,6.0,8.0,10.0\right]$ arcmin. We choose the wavelet scale range as a trade off between the signal-to-noise 
ratio of the string contribution and the small scale foreground contamination. In fact, the wavelet for the smallest scale considered in this analysis peaks at $\ell =1300$, while extending at higher multipoles with a broad distribution. We use maps at an \texttt{HEALPix} resolution of $N_{\rm side} = 2048$, including multipoles till 
$\ell_{\rm max}=2500$. We analyse the simulations with the same U73 mask on the {\it Planck} CMB map (refer to \citealt{planck2013-p06}), which masks both diffuse and compact foregrounds, leaving 73\% of the sky remaining for further analysis (refer to discussion in Sect.~\ref{sec:bispectrumresults}). 

\begin{figure}
\center
\includegraphics[width=8.8cm]{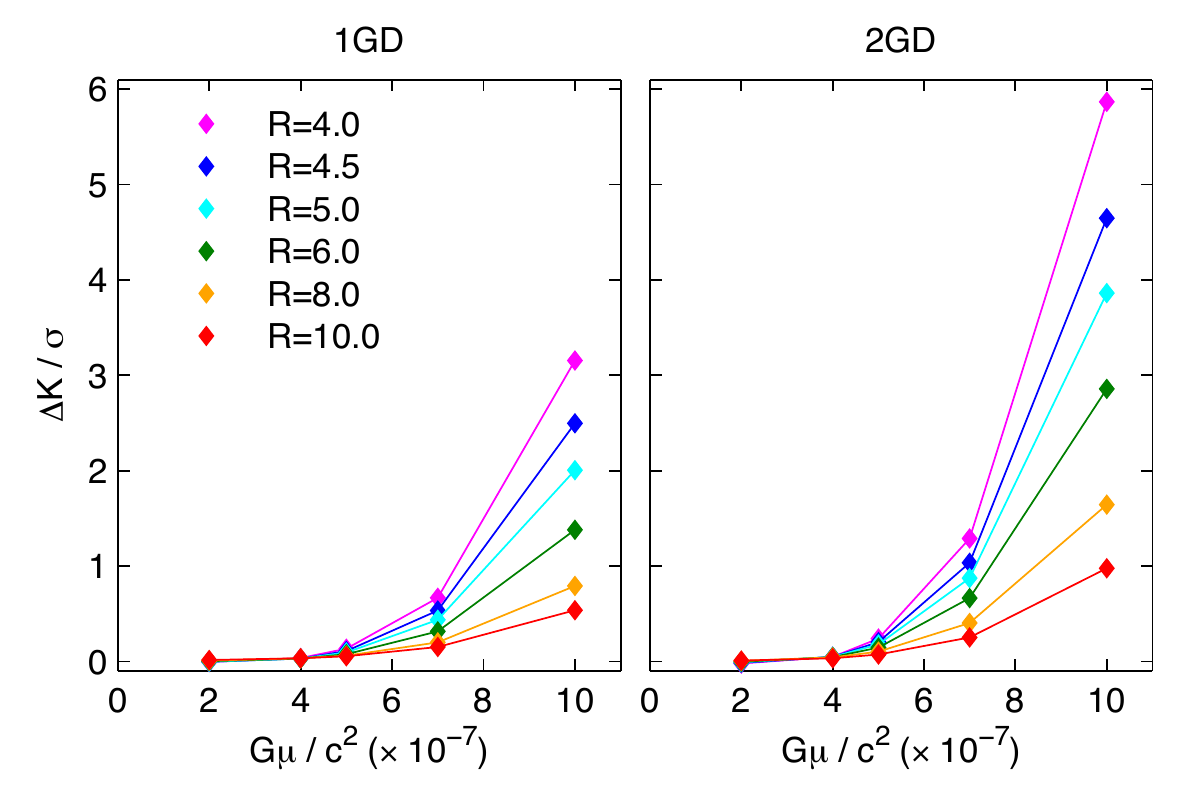}
\caption{Deviation of the kurtosis of the signed-intensity as a function of $\Gmu$, normalized to the standard deviation of CMB and noise simulations: $\Delta K/\sigma = (K(R,\Gmu)-K(R,\Gmu=0))/\sigma$. The left panel shows results for the 1GD wavelet and the right shows the 2GD wavelet. Each curve corresponds to a wavelet scale, $R$ (arcmin), included in the analysis. The final sensitivity of the method is determined by combining the two wavelets and all the scales.}
\label{fig:kurto_rg}
\end{figure}

The string non-Gaussian signatures are characterized in terms of the kurtosis of the signed-intensity $I(\omega_{0},R)$ in \eqref{eq:swr_1} at the different scales $R$ and for both the 1GD and 2GD wavelets. The averaged results from the non-Gaussian simulations were used to model the distribution of the kurtosis as functions of $\Gmu$, i.e., $K(R, \Gmu)$. Other statistics, such as the skewness and the Higher-Criticism, have also been explored. We found that the kurtosis sensitivity to the string tension is higher than the alternative measures. In Fig.~\ref{fig:kurto_rg}, we show the difference between the average kurtosis at several $\Gmu$ values and the average kurtosis for $\Gmu=0$, normalized to the standard deviation of the simulations. On the given range of scales, the 2GD wavelet appears to be more sensitive to the string signal. The final sensitivity of the method in recovering the string tension from simulated data is established through a goodness-of-fit test, performed jointly on the two wavelets for all the scales, and taking into account the correlations by means of a covariance matrix estimated from CMB and noise simulations. From the distribution of the $\Gmu$ values recovered from simulations the null hypothesis can be rejected at  95\% CL for $\Gmu > 7 \times 10^{-7}$. 

The procedure described above was applied to the \textsf{SEVEM} CMB cleaned map, 
observing no evidence of a string signal at any of the scales studied.   This would imply  a wavelet constraint on the string tension at 95\% CL,
$\Gmu < 7 \times 10^{-7}$, however, we note that the signed intensity kurtosis of the data is smaller than that of the model even for small values of $\Gmu$, and it is not fully compatible with it. 
A similar behavior is observed for the skewness of $I\left(\omega_{0},R\right)$.
We evaluated the impact on the estimator of unresolved point sources, both from radio galaxies and sub-millimetre star-forming galaxies, using simulations of these astrophysical components processed through the component separation pipelines \citep{planck2013-p06}. We found that at the smallest wavelet scale considered in this analysis, $R=4$ arcmin, these residual foregrounds induce a shift in the kurtosis, $\Delta K/\sigma = 0.03$, i.e., a bias that is negligible for the present analysis. This shift increases to more than $\Delta K/\sigma = 0.3$ when extending the analysis to wavelet scales as small as $R=2.5$ arcmin. However, the shift induced by unresolved point sources increases the kurtosis of the signed intensity, so it could not reconcile the tension between the model and the data pointed out above. Validation tests performed on the String Challenge data set demonstrated that the sensitivity of the estimator can improve to $\Delta\Gmu  \approx 4 \times 10^{-7}$ for small values of $\Gmu$ when wavelet small scales are included. For this reason, we believe it could be a powerful tool in constraining strings, even though further investigation is needed to reliably modeling the string signal, the foreground contribution and the noise properties, in order to fully exploit the sensitivity of the {\it Planck} data. 

Finally, we note that we have also endeavoured to study the simulated string maps using spherical wavelets, making an extension of previous work \citep{wiaux:2007:sdw,hammond:2009} to compute the Bayesian posterior distribution of the string tension.  Both the spherical and steerable wavelet methods offer good prospects for improved non-Gaussian string constraints from the \Planck\ full mission data.

\subsection{Real space tests for cosmic strings}

\subsubsection{Minkowski functionals method}

Minkowski Functionals (MFs) describe morphological properties of the CMB field, and can be used as generic estimators of non-Gaussianities. They have long been considered to constrain cosmic string physics, for example on gradient temperature maps (see e.g., \citealt{gott1990}). Indeed they have sensitivity to non-Gaussianity sourced by strings at all orders (i.e., including the kurtosis or trispectrum) and they could prove to be a powerful tool to constrain topological defects in general.   For the sake of brevity and conciseness, precise definitions of MFs and analytic formulations are presented in \cite{planck2013-p09} and, here, we only review how MFs can be used to constrain the string energy density $\Gmu$, following the method discussed in \cite{ducout2012}, for the local model $f_{\rm NL}^{\rm local}$.

We measure the four normalized\footnote{Raw Minkowski functionals $V_k$ depend on the Gaussian part of fields through a normalization factor $A_k$, that is a function only of the power spectrum shape. We therefore normalize functionals $v_k=V_k/A_k$ to focus on non-Gaussianity, see \cite{planck2013-p09} and references therein.} functionals $v_k\; (k=0,3)$ (respectively Area, Perimeter, Genus and $N_{\rm cluster}$), computed on $n_{\rm th}=26$ thresholds $\nu$, between $\nu_{\rm min}=-3.5$ and $\nu_{\rm max}=+3.5$ in units of the standard deviation of the map. The curves are combined into one vector $y$ (of size $n=104$).

\begin{figure}[b]
\center
\includegraphics[width=8.8cm]{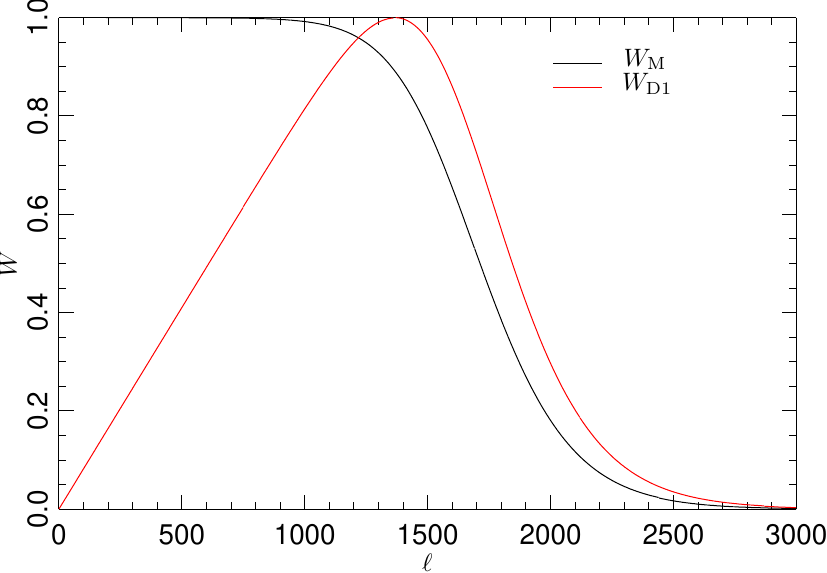}
\caption{The two Wiener filters, $W_{\rm M}$ and $W_{\rm D1}$ used to constrain $\Gmu$ with Minkowski Functionals.}
\label{fig:Wiener_filter_MFs}
\end{figure}

The principle is then to compare measurement of MFs on \Planck\ data  $\hat{y}$ to measurements on NG simulations including a string tension level $\Gmu$, that is, we determine $\hat{y }( \Gmu )$ numerically. These simulations reproduce the same systematics (noise, beam) and the same processing (filter, masks) as the data. Assuming that MFs are multi-variate Gaussians, we obtain a posterior distribution for $\Gmu$ through a $\chi^{2}$ test :
\begin{equation}
 P ( \Gmu | \,\hat{y} ) \propto \exp \left [ -\dfrac{ \chi^{2} (\hat{y},\Gmu) }{2} \right]\,,
\label{eq:postsimp}
\end{equation}
with 
\begin{equation}
\chi^{2}(\hat{y}, \Gmu)  \equiv   \left[ \hat{y}- \bar{y}( \Gmu )   \right]^{T}C^{-1} \left[ \hat{y}- \bar{y}( \Gmu ) \right].
\label{eq:mychi2}
\end{equation}
The covariance matrix $C$ is computed from $10^4$ \textit{Gaussian} simulations\footnote{The Gaussian simulations endeavour to incorporate realistic noise from the {\it Planck} data, but only the effective isotropic beam of the component separation method.} because, given the existing stringent constraints on cosmic strings, this should be accurate without biasing results. 
The cosmic string MF curve $\bar{y}( \Gmu )$ is calibrated on $10^3$ realistic lensed \textit{Planck} simulations, to which we have added a string component at a specified level. These simulations take into account the asymmetry of beams and the component separation process (FFP6 simulations, see \cite{planck2013-p28} for a detailed description). For the string component, we had at our disposal only two high resolution string simulations \citep{2012arXiv1204.5041R}, so our model is the averaged curve obtained from this combination of \textit{Planck} and string simulations.

Due to the nonlinear dependence of MFs on $\Gmu$ and the small number of string simulations, the posterior distribution is quite complex and noisy. For this reason, we evaluated the posterior at $n_{_{\rm NL}}=51$ values of $\Gmu$, between $0$ and $10\times 10^{-7}$, to obtain our \Planck\ estimate for ${\Gmu}$. This estimate is stable and has been validated in realistic conditions with the \Planck\ String Challenges described above, and  for which we found consistent results with the underlying (unknown) $\Gmu$. 
 
\subsubsection{Minkowski functionals results}

For the constraint on $\Gmu$, we analysed the foregrounds separated \texttt{SMICA} map at $N_{\rm side}=2048$ and $\ell_{\rm max}=2000$, using the U73 mask ($f_{\rm sky}=73$\% of the sky is unmasked). The smallest point sources holes were inpainted. 
We applied two specific Wiener filters to the map, designed to enhance the information from the map itself ($W_{\rm M}$) and from the gradients of the map ($W_{\rm D1}=\sqrt{\ell (\ell +1)} W_{\rm M}$). The filters are shown in Fig.~\ref{fig:Wiener_filter_MFs}.

Additionally, we estimated the average impact of some residual foregrounds and secondaries (FG) on ${\Gmu}$, using the linear properties of MFs and foregrounds models processed through the \textit{Planck} simulation pipeline (FFP6 simulations, see \cite{planck2013-p28}). Uncorrelated (Poissonian) unresolved point sources (PS), Cosmic Infrared Background (CIB) and Sunyaev-Zeldovich cluster\footnote{The SZ signal does not include the SZ-lensing NG contribution.} (SZ) signals can be introduced as a simple additive bias $\Delta \bar{y}^{\rm PS, ...}$ on MF curves following:
\begin{equation}
\hat{y}=\hat{y}^{\rm FG subtracted}+\Delta \bar{y}^{\rm PS}+\Delta \bar{y}^{\rm CIB}+\Delta \bar{y}^{\rm SZ}.
\end{equation}
These biases are obtained as an average from 100 simulations, however, these do not comprehensively cover all the different component contributions in the actual \Planck\ data.

\begin{figure}[t!]
\center
\includegraphics[width=8.8cm]{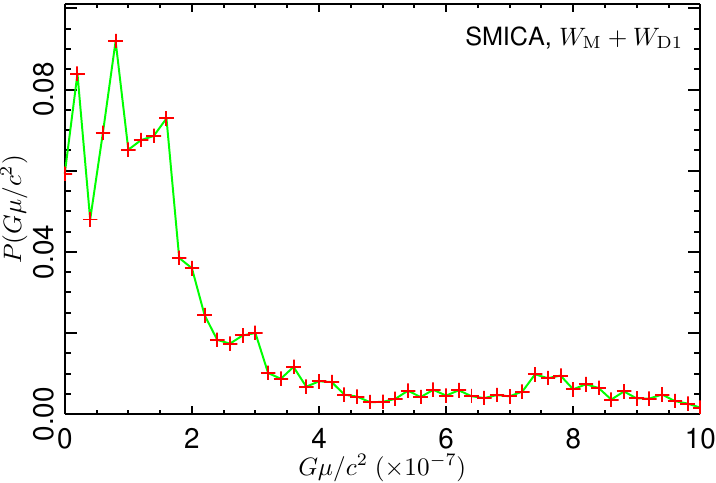}
\caption{Posterior distribution of the parameter $\Gmu$ obtained with Minkowski functionals. This estimate takes into account the lensing of the data, but not the effects of foreground residuals.}
\label{minkowski_f1}
\end{figure}

We eventually obtain the posterior distribution of $\Gmu$, and we integrate it to report confidence intervals. 
Results are summarized in Table~\ref{tab:mfs_results}, for raw data (lensing subtracted) and foreground subtracted data (PS, CIB and SZ subtracted). The discrepancy between the two filters can be explained because the derivative filter $W_{\rm D1}$ scans smaller scales than $W_{\rm M}$ so it is more easily biased by foreground residuals.   Given the remaining foreground uncertainties, we take the most conservative MF constraint for the cosmic string contribution to the \Planck\ data to be
\begin{center}
$\Gmu < 7.8\times 10^{-7}$ \qquad at 95\% C.L. 
\end{center}
The corresponding posterior is presented in Fig.~\ref{minkowski_f1}.

Some caveats need to be mentioned that may influence these results. First, for the MF method itself, an important 
limitation is the small number of string simulations used to calibrate the estimator. The estimator appears to be mostly sensitive to low-redshift strings (infinite strings, with redshifts between 0 and 30), and this is affected by cosmic variance. As low-redshift string simulations are much faster to produce than complete simulations back to recombination, it should be possible to improve the robustness of the constraint using these relatively soon.   Secondly, the impacts of the different point-source foreground components (here, PS, CIB and SZ) have been evaluated by averaging over 100 \Planck\ simulation maps for which the modelling is only partial. The precise contributions of these different components needs to be investigated in more detail for the \Planck\ data. Fortunately, using the linearity of MFs for these contributions it  will be possible to jointly estimate these as their characterisation improves in future studies. Finally, the impact of Galactic residuals should also be assessed in further detail, especially for the filter $W_{\rm D1}$ that we have observed to be less robust against residuals than the $W_{\rm M}$ filter. 

With advances in studying these experimental effects there are good prospects for the full mission data, the sensitivity of the MFs estimator should improve substantially, with simulations forecasting possible MF cosmic string constraints of $\Gmu < 3 \times 10^{-7}$ at the 95\% C.L.   We note that further real space analysis of string map simulations has been undertaken with scaling indices of the pixel temperature distribution (see, e.g., \citealt{2011MNRAS.415.2205R}).  Extensions calculating a set of anisotropic scaling indices along predefined  directions appear to offer good prospects fro string detection.

\begin{table}[tmb]
\caption{MFs constraints obtained on $\Gmu$, at the 95\% C.L. These results are obtained on the \texttt{SMICA} map with the U73 mask ($f_{\rm sky}=0.73$). The ``Raw map'' result includes only the lensing contribution to the data, while the ``Foreground subtracted map'' includes the lensing, Poissonian point sources, CIB and SZ clusters contributions.}
\label{tab:mfs_results}
\begin{center}
\begin{tabular}{l cc}
    \noalign{\doubleline\vskip 2pt}
$\Gmu$ & $W_{\rm M}$ & $W_{\rm M}+W_{\rm D1}$ \\
\noalign{\vskip 3pt\hrule\vskip 4pt}
 Raw map &  $<$ 6.8 $\times 10^{-7}$       & $<$ 7.8 $\times 10^{-7}$\\ 
 FG subtracted map & $<$ 6.0 $\times 10^{-7}$ & $<$ 3.6 $\times 10^{-7}$\\
\noalign{\vskip 3pt\hrule\vskip 4pt}
\end{tabular}
\end{center}
\end{table}

\section{Conclusions}

We have reviewed the signatures induced by cosmic strings in the CMB and searched for these in the \Planck\ data, resulting in new more stringent constraints on the dimensionless string tension parameter $\Gmu$.   A pre-requisite for accurate constraints on cosmic strings is a quantitative understanding of both cosmological string network evolution and the effects they induce in the CMB.   These are  computationally demanding problems but  progress has been made recently on several fronts:   First, high resolution simulations of Nambu-Goto strings have yielded robust results for the scale-invariant properties of string networks on large scales, while there has been increasing convergence about small-scale structure and loops (for which the CMB predictions are less sensitive).  Secondly, post-recombination gravitational effects of strings have been incorporated into full-sky \Planck\ resolution CMB temperature maps  that are important for validating non-Gaussian search methods.   Finally, fast Boltzmann pipelines to calculate CMB power spectra induced by causal sources have been developed and tested at high resolution, whether for field theory simulations of strings or textures or for models of Nambu-Goto strings.  Three-dimensional field theory simulations of vortex-strings at sufficient resolution should, in principle, converge towards the one-dimensional Nambu-Goto string simulations, but testing this is not numerically feasible at present.   For this reason, we believe it is prudent to also include constraints on field theory strings (labelled $G\mu_{\rm AH}$), thus encompassing cosmic string models for which radiative effects are important at late times (such as global strings).   We believe this brackets the important theoretical uncertainties that remain, that is, we have used the best available information to constrain both Nambu-Goto strings (NAMBU) and field theory strings (AH).   This work has also obtained more stringent constraints on semilocal strings and global textures. 

\subsection{Cosmic string constraints and the CMB power spectrum} 

Accurate forecasts for the CMB power spectrum induced by cosmic strings are more difficult to compute than their equivalent for simple adiabatic inflationary scenarios.  It requires knowledge of the source, quantified by the unequal-time correlator (UETC) of the defect stress-energy tensor, from well before recombination to the present day, which is not computationally feasible. Fortunately, we can exploit scale-invariant string evolution to extrapolate the results of simulations with substantially smaller dynamic range. We use two methods to obtain predictions for the UETCs.  An unconnected segment model (USM) is used to model the properties of an evolving string network, determining its density from an analytic one-scale evolution model, and the sources are coupled to the line-of-sight Boltzmann solver  \texttt{CMBACT}.  A second independent pipeline measures the UETCs directly from string simulations in the abelian-Higgs field theory, passing these to a modified form of the \texttt{CMBEASY} Boltzmann code.   The resulting Nambu-Goto and abelian-Higgs string CMB power spectra are illustrated in Fig.~\ref{fig:comp}.   Free parameters in the USM model can be chosen to phenomenologically match the field theory UETCs (denoted the AH-mimic model) and the comparison is also shown in Fig.~\ref{fig:comp}, validating the two independent pipelines. 

To compute constraints on cosmic string scenarios we have added the angular power spectrum to that for a simple adiabatic model, assuming that they are uncorrelated, with the fraction of the spectrum contributed by cosmic strings being $f_{10}$ at $\ell=10$. This has been added to the standard 6 parameter fit using {\tt COSMOMC} with flat priors.  For the USM models we have obtained the constraint for the Nambu-Goto string model
\eq
 G\mu/c^2 < 1.5\times 10^{-7}\,, \qquad \qquad f_{10} < 0.015\,,
\qe
while for the abelian-Higgs field theory model we find,
\eq
G\mu_{\rm AH}/c^2 < 3.2\times 10^{-7}\,, \qquad \qquad f_{10} < 0.028\,. 
\qe
 The  marginalized likelihoods for $f_{10}$  and in the $f_{10}$--$\Omega_{\mathrm b}h^2$ plane were presented in Fig.~\ref{fig:marginalized}. 
 With \Planck\ nominal mission data these limits are already about a factor of two more stringent than the comparable \textit{WMAP} 7-year string constraints and these \Planck\ limits improve further with the inclusion of high-$\ell$ data.   

\subsection{Non-Gaussian searches for cosmic strings}

Complementary searches for non-Gaussian signatures from cosmic strings were performed and we have reported constraints from the string bispectrum, steerable wavelets and Minkowski functionals.  These methods participated in the \Planck\ String Challenges and have undergone non-Gaussian validation tests.   

The post-recombination string bispectrum has been reconstructed and calibrated from string-induced CMB maps using a modal estimator.
String Challenge analysis with \Planck-realistic noise simulations and mask indicated a nominal mission sensitivity of $\Delta \Gmu  \approx 5.8 \times 10^{-7}$.
Analysis of \texttt{SMICA}, \texttt{NILC} and \texttt{SEVEM} foreground-separated maps has yielded $f_{\rm NL} = 0.37\pm 0.21$ for the string bispectrum shape, which translates into a bispectrum constraint on the string tension,
$\Gmu< 8.8\times 10^{-7}$  (95 \% CL).    Steerable wavelet methods have been calibrated on string simulation maps added to Gaussian CMB maps with realistic noise and masking, showing a sensitivity of up to $\Delta \Gmu  \approx 4 \times 10^{-7}$.  The string signal was shown to have greater impact on the kurtosis  of the signed-intensity than on its skewness, and no evidence of a string signal was found in the \Planck\ data.    Minkowski functionals have been applied to string simulation maps in a \Planck-realistic context, computing the four functionals---area, perimeter, genus and $N_{\rm cluster}$---after application of Weiner filters.  Using these distributions, a Bayesian estimator has been constructed to constrain the string tension.    
Analysis of the \texttt{SMICA} foreground-cleaned maps yielded a MF constraint of 
$\Gmu< 7.8\times 10^{-7}$  (95 \% CL).  

Non-Gaussian searches for strings are complementary to the power spectrum analysis and yield constraints as low as  $\Gmu < 7.8 \times10^{-7}$, though we note the potential impact of foreground residuals in limiting current precision. These are conservative upper bounds because they only include post-recombination string contributions, unlike the string power spectrum analysis.  Having such a broad suite of tools, ranging from multipole space, through wavelets, to real space detection methods, allows cross-validation and  reinforces the conclusion that there is at present no evidence for cosmic strings in the \Planck\  nominal mission data.

\begin{acknowledgements}

The development of \Planck\ has been supported by: ESA; CNES and CNRS/INSU-IN2P3-INP (France); ASI, CNR, and INAF (Italy); NASA and DoE (USA); STFC and UKSA (UK); CSIC, MICINN, JA and RES (Spain); Tekes, AoF and CSC (Finland); DLR and MPG (Germany); CSA (Canada); DTU Space (Denmark); SER/SSO (Switzerland); RCN (Norway); SFI (Ireland); FCT/MCTES (Portugal); and PRACE (EU). A description of the \Planck\ Collaboration and a list of its members, including the technical or scientific activities in which they have been involved, can be found at \url{http://www.sciops.esa.int/index.php?project=planck&page=Planck_Collaboration}.   We also wish to acknowledge the use of the COSMOS supercomputer, part of the DiRAC HPC Facility funded by STFC and the UK Large Facilities Capital Fund, as well as the use of the Andromeda cluster of the University of Geneva.

\end{acknowledgements}

\bibliography{Defects,Planck_bib}{}

\raggedright

\end{document}

%% file: Planck.tex
\def\setsymbol#1#2{\expandafter\def\csname #1\endcsname{#2}}
\def\getsymbol#1{\csname #1\endcsname}

\def\Planck{\textit{Planck}}

\newbox\tablebox    \newdimen\tablewidth
\def\leaderfil{\leaders\hbox to 5pt{\hss.\hss}\hfil}
\def\endPlancktablewide{\tablewidth=\textwidth 
    $$\hss\copy\tablebox\hss$$
    \vskip-\lastskip\vskip -2pt}
\def\tablenote#1 #2\par{\begingroup \parindent=0.8em
    \abovedisplayshortskip=0pt\belowdisplayshortskip=0pt
    \noindent
    $$\hss\vbox{\hsize\tablewidth \hangindent=\parindent \hangafter=1 \noindent
    \hbox to \parindent{$^#1$\hss}\strut#2\strut\par}\hss$$
    \endgroup}
\def\doubleline{\vskip 3pt\hrule \vskip 1.5pt \hrule \vskip 5pt}

\def\L2{\ifmmode L_2\else $L_2$\fi}

\def\DeltaT{\ifmmode \Delta T\else $\Delta T$\fi}
\def\deltat{\ifmmode \Delta t\else $\Delta t$\fi}
\def\fknee{\ifmmode f_{\rm knee}\else $f_{\rm knee}$\fi}
\def\Fmax{\ifmmode F_{\rm max}\else $F_{\rm max}$\fi}
\def\solar{\ifmmode{\rm M}_{\mathord\odot}\else${\rm M}_{\mathord\odot}$\fi}
\def\Msolar{\ifmmode{\rm M}_{\mathord\odot}\else${\rm M}_{\mathord\odot}$\fi}
\def\Lsolar{\ifmmode{\rm L}_{\mathord\odot}\else${\rm L}_{\mathord\odot}$\fi}

\def\inv{\ifmmode^{-1}\else$^{-1}$\fi}
\def\mo{\ifmmode^{-1}\else$^{-1}$\fi}
\def\sup#1{\ifmmode ^{\rm #1}\else $^{\rm #1}$\fi}
\def\expo#1{\ifmmode \times 10^{#1}\else $\times 10^{#1}$\fi}
\def\,{\thinspace}
\def\lsim{\mathrel{\raise .4ex\hbox{\rlap{$<$}\lower 1.2ex\hbox{$\sim$}}}}
\def\gsim{\mathrel{\raise .4ex\hbox{\rlap{$>$}\lower 1.2ex\hbox{$\sim$}}}}

\def\simprop{\mathrel{\raise .4ex\hbox{\rlap{$\propto$}\lower 1.2ex\hbox{$\sim$}}}}
\def\deg{\ifmmode^\circ\else$^\circ$\fi}
\def\pdeg{\ifmmode $\setbox0=\hbox{$^{\circ}$}\rlap{\hskip.11\wd0 .}$^{\circ}
          \else \setbox0=\hbox{$^{\circ}$}\rlap{\hskip.11\wd0 .}$^{\circ}$\fi}
\def\arcs{\ifmmode {^{\scriptstyle\prime\prime}}
          \else $^{\scriptstyle\prime\prime}$\fi}
\def\arcm{\ifmmode {^{\scriptstyle\prime}}
          \else $^{\scriptstyle\prime}$\fi}
\newdimen\sa  \newdimen\sb
\def\parcs{\sa=.07em \sb=.03em
     \ifmmode \hbox{\rlap{.}}^{\scriptstyle\prime\kern -\sb\prime}\hbox{\kern -\sa}
     \else \rlap{.}$^{\scriptstyle\prime\kern -\sb\prime}$\kern -\sa\fi}
\def\parcm{\sa=.08em \sb=.03em
     \ifmmode \hbox{\rlap{.}\kern\sa}^{\scriptstyle\prime}\hbox{\kern-\sb}
     \else \rlap{.}\kern\sa$^{\scriptstyle\prime}$\kern-\sb\fi}
\def\ra[#1 #2 #3.#4]{#1\sup{h}#2\sup{m}#3\sup{s}\llap.#4}
\def\dec[#1 #2 #3.#4]{#1\deg#2\arcm#3\arcs\llap.#4}
\def\deco[#1 #2 #3]{#1\deg#2\arcm#3\arcs}
\def\rra[#1 #2]{#1\sup{h}#2\sup{m}}

\def\dots{\relax\ifmmode \ldots\else $\ldots$\fi}
\def\WHzsr{\ifmmode $W\,Hz\mo\,sr\mo$\else W\,Hz\mo\,sr\mo\fi}
\def\mHz{\ifmmode $\,mHz$\else \,mHz\fi}
\def\GHz{\ifmmode $\,GHz$\else \,GHz\fi}
\def\mKs{\ifmmode $\,mK\,s$^{1/2}\else \,mK\,s$^{1/2}$\fi}
\def\muKs{\ifmmode \,\mu$K\,s$^{1/2}\else \,$\mu$K\,s$^{1/2}$\fi}
\def\muKRJs{\ifmmode \,\mu$K$_{\rm RJ}$\,s$^{1/2}\else \,$\mu$K$_{\rm RJ}$\,s$^{1/2}$\fi}
\def\muKHz{\ifmmode \,\mu$K\,Hz$^{-1/2}\else \,$\mu$K\,Hz$^{-1/2}$\fi}
\def\MJysr{\ifmmode \,$MJy\,sr\mo$\else \,MJy\,sr\mo\fi}
\def\MJysrmK{\ifmmode \,$MJy\,sr\mo$\,mK$_{\rm CMB}\mo\else \,MJy\,sr\mo\,mK$_{\rm CMB}\mo$\fi}
\def\microns{\ifmmode \,\mu$m$\else \,$\mu$m\fi}

\def\muK{\ifmmode \,\mu$K$\else \,$\mu$\hbox{K}\fi}
\def\microK{\ifmmode \,\mu$K$\else \,$\mu$\hbox{K}\fi}
\def\muW{\ifmmode \,\mu$W$\else \,$\mu$\hbox{W}\fi}
\def\kms{\ifmmode $\,km\,s$^{-1}\else \,km\,s$^{-1}$\fi}
\def\kmsMpc{\ifmmode $\,\kms\,Mpc\mo$\else \,\kms\,Mpc\mo\fi}
\setsymbol{LFI:center:frequency:70GHz:units}{70.3\,GHz}
\setsymbol{LFI:center:frequency:44GHz:units}{44.1\,GHz}
\setsymbol{LFI:center:frequency:30GHz:units}{28.5\,GHz}

\setsymbol{LFI:center:frequency:70GHz}{70.3}
\setsymbol{LFI:center:frequency:44GHz}{44.1}
\setsymbol{LFI:center:frequency:30GHz}{28.5}

\setsymbol{LFI:center:frequency:LFI18:Rad:M:units}{71.7\GHz}
\setsymbol{LFI:center:frequency:LFI19:Rad:M:units}{67.5\GHz}
\setsymbol{LFI:center:frequency:LFI20:Rad:M:units}{69.2\GHz}
\setsymbol{LFI:center:frequency:LFI21:Rad:M:units}{70.4\GHz}
\setsymbol{LFI:center:frequency:LFI22:Rad:M:units}{71.5\GHz}
\setsymbol{LFI:center:frequency:LFI23:Rad:M:units}{70.8\GHz}
\setsymbol{LFI:center:frequency:LFI24:Rad:M:units}{44.4\GHz}
\setsymbol{LFI:center:frequency:LFI25:Rad:M:units}{44.0\GHz}
\setsymbol{LFI:center:frequency:LFI26:Rad:M:units}{43.9\GHz}
\setsymbol{LFI:center:frequency:LFI27:Rad:M:units}{28.3\GHz}
\setsymbol{LFI:center:frequency:LFI28:Rad:M:units}{28.8\GHz}
\setsymbol{LFI:center:frequency:LFI18:Rad:S:units}{70.1\GHz}
\setsymbol{LFI:center:frequency:LFI19:Rad:S:units}{69.6\GHz}
\setsymbol{LFI:center:frequency:LFI20:Rad:S:units}{69.5\GHz}
\setsymbol{LFI:center:frequency:LFI21:Rad:S:units}{69.5\GHz}
\setsymbol{LFI:center:frequency:LFI22:Rad:S:units}{72.8\GHz}
\setsymbol{LFI:center:frequency:LFI23:Rad:S:units}{71.3\GHz}
\setsymbol{LFI:center:frequency:LFI24:Rad:S:units}{44.1\GHz}
\setsymbol{LFI:center:frequency:LFI25:Rad:S:units}{44.1\GHz}
\setsymbol{LFI:center:frequency:LFI26:Rad:S:units}{44.1\GHz}
\setsymbol{LFI:center:frequency:LFI27:Rad:S:units}{28.5\GHz}
\setsymbol{LFI:center:frequency:LFI28:Rad:S:units}{28.2\GHz}

\setsymbol{LFI:center:frequency:LFI18:Rad:M}{71.7}
\setsymbol{LFI:center:frequency:LFI19:Rad:M}{67.5}
\setsymbol{LFI:center:frequency:LFI20:Rad:M}{69.2}
\setsymbol{LFI:center:frequency:LFI21:Rad:M}{70.4}
\setsymbol{LFI:center:frequency:LFI22:Rad:M}{71.5}
\setsymbol{LFI:center:frequency:LFI23:Rad:M}{70.8}
\setsymbol{LFI:center:frequency:LFI24:Rad:M}{44.4}
\setsymbol{LFI:center:frequency:LFI25:Rad:M}{44.0}
\setsymbol{LFI:center:frequency:LFI26:Rad:M}{43.9}
\setsymbol{LFI:center:frequency:LFI27:Rad:M}{28.3}
\setsymbol{LFI:center:frequency:LFI28:Rad:M}{28.8}
\setsymbol{LFI:center:frequency:LFI18:Rad:S}{70.1}
\setsymbol{LFI:center:frequency:LFI19:Rad:S}{69.6}
\setsymbol{LFI:center:frequency:LFI20:Rad:S}{69.5}
\setsymbol{LFI:center:frequency:LFI21:Rad:S}{69.5}
\setsymbol{LFI:center:frequency:LFI22:Rad:S}{72.8}
\setsymbol{LFI:center:frequency:LFI23:Rad:S}{71.3}
\setsymbol{LFI:center:frequency:LFI24:Rad:S}{44.1}
\setsymbol{LFI:center:frequency:LFI25:Rad:S}{44.1}
\setsymbol{LFI:center:frequency:LFI26:Rad:S}{44.1}
\setsymbol{LFI:center:frequency:LFI27:Rad:S}{28.5}
\setsymbol{LFI:center:frequency:LFI28:Rad:S}{28.2}

\setsymbol{LFI:white:noise:sensitivity:70GHz:units}{134.7\muKs}
\setsymbol{LFI:white:noise:sensitivity:44GHz:units}{164.7\muKs}
\setsymbol{LFI:white:noise:sensitivity:30GHz:units}{143.4\muKs}

\setsymbol{LFI:white:noise:sensitivity:70GHz}{134.7}
\setsymbol{LFI:white:noise:sensitivity:44GHz}{164.7}
\setsymbol{LFI:white:noise:sensitivity:30GHz}{143.4}

\setsymbol{LFI:white:noise:sensitivity:LFI18:Rad:M:units}{512.0\muKs}
\setsymbol{LFI:white:noise:sensitivity:LFI19:Rad:M:units}{581.4\muKs}
\setsymbol{LFI:white:noise:sensitivity:LFI20:Rad:M:units}{590.8\muKs}
\setsymbol{LFI:white:noise:sensitivity:LFI21:Rad:M:units}{455.2\muKs}
\setsymbol{LFI:white:noise:sensitivity:LFI22:Rad:M:units}{492.0\muKs}
\setsymbol{LFI:white:noise:sensitivity:LFI23:Rad:M:units}{507.7\muKs}
\setsymbol{LFI:white:noise:sensitivity:LFI24:Rad:M:units}{462.2\muKs}
\setsymbol{LFI:white:noise:sensitivity:LFI25:Rad:M:units}{413.6\muKs}
\setsymbol{LFI:white:noise:sensitivity:LFI26:Rad:M:units}{478.6\muKs}
\setsymbol{LFI:white:noise:sensitivity:LFI27:Rad:M:units}{277.7\muKs}
\setsymbol{LFI:white:noise:sensitivity:LFI28:Rad:M:units}{312.3\muKs}
\setsymbol{LFI:white:noise:sensitivity:LFI18:Rad:S:units}{465.7\muKs}
\setsymbol{LFI:white:noise:sensitivity:LFI19:Rad:S:units}{555.6\muKs}
\setsymbol{LFI:white:noise:sensitivity:LFI20:Rad:S:units}{623.2\muKs}
\setsymbol{LFI:white:noise:sensitivity:LFI21:Rad:S:units}{564.1\muKs}
\setsymbol{LFI:white:noise:sensitivity:LFI22:Rad:S:units}{534.4\muKs}
\setsymbol{LFI:white:noise:sensitivity:LFI23:Rad:S:units}{542.4\muKs}
\setsymbol{LFI:white:noise:sensitivity:LFI24:Rad:S:units}{399.2\muKs}
\setsymbol{LFI:white:noise:sensitivity:LFI25:Rad:S:units}{392.6\muKs}
\setsymbol{LFI:white:noise:sensitivity:LFI26:Rad:S:units}{418.6\muKs}
\setsymbol{LFI:white:noise:sensitivity:LFI27:Rad:S:units}{302.9\muKs}
\setsymbol{LFI:white:noise:sensitivity:LFI28:Rad:S:units}{285.3\muKs}

\setsymbol{LFI:white:noise:sensitivity:LFI18:Rad:M}{512.0}
\setsymbol{LFI:white:noise:sensitivity:LFI19:Rad:M}{581.4}
\setsymbol{LFI:white:noise:sensitivity:LFI20:Rad:M}{590.8}
\setsymbol{LFI:white:noise:sensitivity:LFI21:Rad:M}{455.2}
\setsymbol{LFI:white:noise:sensitivity:LFI22:Rad:M}{492.0}
\setsymbol{LFI:white:noise:sensitivity:LFI23:Rad:M}{507.7}
\setsymbol{LFI:white:noise:sensitivity:LFI24:Rad:M}{462.2}
\setsymbol{LFI:white:noise:sensitivity:LFI25:Rad:M}{413.6}
\setsymbol{LFI:white:noise:sensitivity:LFI26:Rad:M}{478.6}
\setsymbol{LFI:white:noise:sensitivity:LFI27:Rad:M}{277.7}
\setsymbol{LFI:white:noise:sensitivity:LFI28:Rad:M}{312.3}
\setsymbol{LFI:white:noise:sensitivity:LFI18:Rad:S}{465.7}
\setsymbol{LFI:white:noise:sensitivity:LFI19:Rad:S}{555.6}
\setsymbol{LFI:white:noise:sensitivity:LFI20:Rad:S}{623.2}
\setsymbol{LFI:white:noise:sensitivity:LFI21:Rad:S}{564.1}
\setsymbol{LFI:white:noise:sensitivity:LFI22:Rad:S}{534.4}
\setsymbol{LFI:white:noise:sensitivity:LFI23:Rad:S}{542.4}
\setsymbol{LFI:white:noise:sensitivity:LFI24:Rad:S}{399.2}
\setsymbol{LFI:white:noise:sensitivity:LFI25:Rad:S}{392.6}
\setsymbol{LFI:white:noise:sensitivity:LFI26:Rad:S}{418.6}
\setsymbol{LFI:white:noise:sensitivity:LFI27:Rad:S}{302.9}
\setsymbol{LFI:white:noise:sensitivity:LFI28:Rad:S}{285.3}

\setsymbol{LFI:knee:frequency:70GHz:units}{29.5\mHz}
\setsymbol{LFI:knee:frequency:44GHz:units}{56.2\mHz}
\setsymbol{LFI:knee:frequency:30GHz:units}{113.7\mHz}

\setsymbol{LFI:knee:frequency:70GHz}{29.5}
\setsymbol{LFI:knee:frequency:44GHz}{56.2}
\setsymbol{LFI:knee:frequency:30GHz}{113.7}

\setsymbol{LFI:knee:frequency:LFI18:Rad:M:units}{16.3\mHz}
\setsymbol{LFI:knee:frequency:LFI19:Rad:M:units}{15.1\mHz}
\setsymbol{LFI:knee:frequency:LFI20:Rad:M:units}{18.7\mHz}
\setsymbol{LFI:knee:frequency:LFI21:Rad:M:units}{37.2\mHz}
\setsymbol{LFI:knee:frequency:LFI22:Rad:M:units}{12.7\mHz}
\setsymbol{LFI:knee:frequency:LFI23:Rad:M:units}{34.6\mHz}
\setsymbol{LFI:knee:frequency:LFI24:Rad:M:units}{46.2\mHz}
\setsymbol{LFI:knee:frequency:LFI25:Rad:M:units}{24.9\mHz}
\setsymbol{LFI:knee:frequency:LFI26:Rad:M:units}{67.6\mHz}
\setsymbol{LFI:knee:frequency:LFI27:Rad:M:units}{187.4\mHz}
\setsymbol{LFI:knee:frequency:LFI28:Rad:M:units}{122.2\mHz}
\setsymbol{LFI:knee:frequency:LFI18:Rad:S:units}{17.7\mHz}
\setsymbol{LFI:knee:frequency:LFI19:Rad:S:units}{22.0\mHz}
\setsymbol{LFI:knee:frequency:LFI20:Rad:S:units}{8.7\mHz}
\setsymbol{LFI:knee:frequency:LFI21:Rad:S:units}{25.9\mHz}
\setsymbol{LFI:knee:frequency:LFI22:Rad:S:units}{15.8\mHz}
\setsymbol{LFI:knee:frequency:LFI23:Rad:S:units}{129.8\mHz}
\setsymbol{LFI:knee:frequency:LFI24:Rad:S:units}{100.9\mHz}
\setsymbol{LFI:knee:frequency:LFI25:Rad:S:units}{38.9\mHz}
\setsymbol{LFI:knee:frequency:LFI26:Rad:S:units}{58.9\mHz}
\setsymbol{LFI:knee:frequency:LFI27:Rad:S:units}{104.4\mHz}
\setsymbol{LFI:knee:frequency:LFI28:Rad:S:units}{40.7\mHz}

\setsymbol{LFI:knee:frequency:LFI18:Rad:M}{16.3}
\setsymbol{LFI:knee:frequency:LFI19:Rad:M}{15.1}
\setsymbol{LFI:knee:frequency:LFI20:Rad:M}{18.7}
\setsymbol{LFI:knee:frequency:LFI21:Rad:M}{37.2}
\setsymbol{LFI:knee:frequency:LFI22:Rad:M}{12.7}
\setsymbol{LFI:knee:frequency:LFI23:Rad:M}{34.6}
\setsymbol{LFI:knee:frequency:LFI24:Rad:M}{46.2}
\setsymbol{LFI:knee:frequency:LFI25:Rad:M}{24.9}
\setsymbol{LFI:knee:frequency:LFI26:Rad:M}{67.6}
\setsymbol{LFI:knee:frequency:LFI27:Rad:M}{187.4}
\setsymbol{LFI:knee:frequency:LFI28:Rad:M}{122.2}
\setsymbol{LFI:knee:frequency:LFI18:Rad:S}{17.7}
\setsymbol{LFI:knee:frequency:LFI19:Rad:S}{22.0}
\setsymbol{LFI:knee:frequency:LFI20:Rad:S}{8.7}
\setsymbol{LFI:knee:frequency:LFI21:Rad:S}{25.9}
\setsymbol{LFI:knee:frequency:LFI22:Rad:S}{15.8}
\setsymbol{LFI:knee:frequency:LFI23:Rad:S}{129.8}
\setsymbol{LFI:knee:frequency:LFI24:Rad:S}{100.9}
\setsymbol{LFI:knee:frequency:LFI25:Rad:S}{38.9}
\setsymbol{LFI:knee:frequency:LFI26:Rad:S}{58.9}
\setsymbol{LFI:knee:frequency:LFI27:Rad:S}{104.4}
\setsymbol{LFI:knee:frequency:LFI28:Rad:S}{40.7}

\setsymbol{LFI:slope:70GHz:units}{$-1.03$\mHz}
\setsymbol{LFI:slope:44GHz:units}{$-0.89$\mHz}
\setsymbol{LFI:slope:30GHz:units}{$-0.87$\mHz}

\setsymbol{LFI:slope:70GHz}{$-1.03$}
\setsymbol{LFI:slope:44GHz}{$-0.89$}
\setsymbol{LFI:slope:30GHz}{$-0.87$}

\setsymbol{LFI:slope:LFI18:Rad:M:units}{$-1.04$\mHz}
\setsymbol{LFI:slope:LFI19:Rad:M:units}{$-1.09$\mHz}
\setsymbol{LFI:slope:LFI20:Rad:M:units}{$-0.69$\mHz}
\setsymbol{LFI:slope:LFI21:Rad:M:units}{$-1.56$\mHz}
\setsymbol{LFI:slope:LFI22:Rad:M:units}{$-1.01$\mHz}
\setsymbol{LFI:slope:LFI23:Rad:M:units}{$-0.96$\mHz}
\setsymbol{LFI:slope:LFI24:Rad:M:units}{$-0.83$\mHz}
\setsymbol{LFI:slope:LFI25:Rad:M:units}{$-0.91$\mHz}
\setsymbol{LFI:slope:LFI26:Rad:M:units}{$-0.95$\mHz}
\setsymbol{LFI:slope:LFI27:Rad:M:units}{$-0.87$\mHz}
\setsymbol{LFI:slope:LFI28:Rad:M:units}{$-0.88$\mHz}
\setsymbol{LFI:slope:LFI18:Rad:S:units}{$-1.15$\mHz}
\setsymbol{LFI:slope:LFI19:Rad:S:units}{$-1.00$\mHz}
\setsymbol{LFI:slope:LFI20:Rad:S:units}{$-0.95$\mHz}
\setsymbol{LFI:slope:LFI21:Rad:S:units}{$-0.92$\mHz}
\setsymbol{LFI:slope:LFI22:Rad:S:units}{$-1.01$\mHz}
\setsymbol{LFI:slope:LFI23:Rad:S:units}{$-0.95$\mHz}
\setsymbol{LFI:slope:LFI24:Rad:S:units}{$-0.73$\mHz}
\setsymbol{LFI:slope:LFI25:Rad:S:units}{$-1.16$\mHz}
\setsymbol{LFI:slope:LFI26:Rad:S:units}{$-0.79$\mHz}
\setsymbol{LFI:slope:LFI27:Rad:S:units}{$-0.82$\mHz}
\setsymbol{LFI:slope:LFI28:Rad:S:units}{$-0.91$\mHz}

\setsymbol{LFI:slope:LFI18:Rad:M}{$-1.04$}
\setsymbol{LFI:slope:LFI19:Rad:M}{$-1.09$}
\setsymbol{LFI:slope:LFI20:Rad:M}{$-0.69$}
\setsymbol{LFI:slope:LFI21:Rad:M}{$-1.56$}
\setsymbol{LFI:slope:LFI22:Rad:M}{$-1.01$}
\setsymbol{LFI:slope:LFI23:Rad:M}{$-0.96$}
\setsymbol{LFI:slope:LFI24:Rad:M}{$-0.83$}
\setsymbol{LFI:slope:LFI25:Rad:M}{$-0.91$}
\setsymbol{LFI:slope:LFI26:Rad:M}{$-0.95$}
\setsymbol{LFI:slope:LFI27:Rad:M}{$-0.87$}
\setsymbol{LFI:slope:LFI28:Rad:M}{$-0.88$}
\setsymbol{LFI:slope:LFI18:Rad:S}{$-1.15$}
\setsymbol{LFI:slope:LFI19:Rad:S}{$-1.00$}
\setsymbol{LFI:slope:LFI20:Rad:S}{$-0.95$}
\setsymbol{LFI:slope:LFI21:Rad:S}{$-0.92$}
\setsymbol{LFI:slope:LFI22:Rad:S}{$-1.01$}
\setsymbol{LFI:slope:LFI23:Rad:S}{$-0.95$}
\setsymbol{LFI:slope:LFI24:Rad:S}{$-0.73$}
\setsymbol{LFI:slope:LFI25:Rad:S}{$-1.16$}
\setsymbol{LFI:slope:LFI26:Rad:S}{$-0.79$}
\setsymbol{LFI:slope:LFI27:Rad:S}{$-0.82$}
\setsymbol{LFI:slope:LFI28:Rad:S}{$-0.91$}

\setsymbol{LFI:FWHM:70GHz:units}{13\parcm01}
\setsymbol{LFI:FWHM:44GHz:units}{27\parcm92}
\setsymbol{LFI:FWHM:30GHz:units}{32\parcm65}

\setsymbol{LFI:FWHM:70GHz}{13.01}
\setsymbol{LFI:FWHM:44GHz}{27.92}
\setsymbol{LFI:FWHM:30GHz}{32.65}

\setsymbol{LFI:FWHM:LFI18:units}{13\parcm39}
\setsymbol{LFI:FWHM:LFI19:units}{13\parcm01}
\setsymbol{LFI:FWHM:LFI20:units}{12\parcm75}
\setsymbol{LFI:FWHM:LFI21:units}{12\parcm74}
\setsymbol{LFI:FWHM:LFI22:units}{12\parcm87}
\setsymbol{LFI:FWHM:LFI23:units}{13\parcm27}
\setsymbol{LFI:FWHM:LFI24:units}{22\parcm98}
\setsymbol{LFI:FWHM:LFI25:units}{30\parcm46}
\setsymbol{LFI:FWHM:LFI26:units}{30\parcm31}
\setsymbol{LFI:FWHM:LFI27:units}{32\parcm65}
\setsymbol{LFI:FWHM:LFI28:units}{32\parcm66}

\setsymbol{LFI:FWHM:LFI18}{13.39}
\setsymbol{LFI:FWHM:LFI19}{13.01}
\setsymbol{LFI:FWHM:LFI20}{12.75}
\setsymbol{LFI:FWHM:LFI21}{12.74}
\setsymbol{LFI:FWHM:LFI22}{12.87}
\setsymbol{LFI:FWHM:LFI23}{13.27}
\setsymbol{LFI:FWHM:LFI24}{22.98}
\setsymbol{LFI:FWHM:LFI25}{30.46}
\setsymbol{LFI:FWHM:LFI26}{30.31}
\setsymbol{LFI:FWHM:LFI27}{32.65}
\setsymbol{LFI:FWHM:LFI28}{32.66}

\setsymbol{LFI:FWHM:uncertainty:LFI18:units}{0.170\arcm}
\setsymbol{LFI:FWHM:uncertainty:LFI19:units}{0.174\arcm}
\setsymbol{LFI:FWHM:uncertainty:LFI20:units}{0.170\arcm}
\setsymbol{LFI:FWHM:uncertainty:LFI21:units}{0.156\arcm}
\setsymbol{LFI:FWHM:uncertainty:LFI22:units}{0.164\arcm}
\setsymbol{LFI:FWHM:uncertainty:LFI23:units}{0.171\arcm}
\setsymbol{LFI:FWHM:uncertainty:LFI24:units}{0.652\arcm}
\setsymbol{LFI:FWHM:uncertainty:LFI25:units}{1.075\arcm}
\setsymbol{LFI:FWHM:uncertainty:LFI26:units}{1.131\arcm}
\setsymbol{LFI:FWHM:uncertainty:LFI27:units}{1.266\arcm}
\setsymbol{LFI:FWHM:uncertainty:LFI28:units}{1.287\arcm}

\setsymbol{LFI:FWHM:uncertainty:LFI18}{0.170}
\setsymbol{LFI:FWHM:uncertainty:LFI19}{0.174}
\setsymbol{LFI:FWHM:uncertainty:LFI20}{0.170}
\setsymbol{LFI:FWHM:uncertainty:LFI21}{0.156}
\setsymbol{LFI:FWHM:uncertainty:LFI22}{0.164}
\setsymbol{LFI:FWHM:uncertainty:LFI23}{0.171}
\setsymbol{LFI:FWHM:uncertainty:LFI24}{0.652}
\setsymbol{LFI:FWHM:uncertainty:LFI25}{1.075}
\setsymbol{LFI:FWHM:uncertainty:LFI26}{1.131}
\setsymbol{LFI:FWHM:uncertainty:LFI27}{1.266}
\setsymbol{LFI:FWHM:uncertainty:LFI28}{1.287}

\setsymbol{HFI:center:frequency:100GHz:units}{100\,GHz}
\setsymbol{HFI:center:frequency:143GHz:units}{143\,GHz}
\setsymbol{HFI:center:frequency:217GHz:units}{217\,GHz}
\setsymbol{HFI:center:frequency:353GHz:units}{353\,GHz}
\setsymbol{HFI:center:frequency:545GHz:units}{545\,GHz}
\setsymbol{HFI:center:frequency:857GHz:units}{857\,GHz}

\setsymbol{HFI:center:frequency:100GHz}{100}
\setsymbol{HFI:center:frequency:143GHz}{143}
\setsymbol{HFI:center:frequency:217GHz}{217}
\setsymbol{HFI:center:frequency:353GHz}{353}
\setsymbol{HFI:center:frequency:545GHz}{545}
\setsymbol{HFI:center:frequency:857GHz}{857}

\setsymbol{HFI:Ndetectors:100GHz}{8}
\setsymbol{HFI:Ndetectors:143GHz}{11}
\setsymbol{HFI:Ndetectors:217GHz}{12}
\setsymbol{HFI:Ndetectors:353GHz}{12}
\setsymbol{HFI:Ndetectors:545GHz}{3}
\setsymbol{HFI:Ndetectors:857GHz}{4}

\setsymbol{HFI:FWHM:Maps:100GHz:units}{9\parcm88}
\setsymbol{HFI:FWHM:Maps:143GHz:units}{7\parcm18}
\setsymbol{HFI:FWHM:Maps:217GHz:units}{4\parcm87}
\setsymbol{HFI:FWHM:Maps:353GHz:units}{4\parcm65}
\setsymbol{HFI:FWHM:Maps:545GHz:units}{4\parcm72}
\setsymbol{HFI:FWHM:Maps:857GHz:units}{4\parcm39}
\setsymbol{HFI:FWHM:Maps:100GHz}{9.88}
\setsymbol{HFI:FWHM:Maps:143GHz}{7.18}
\setsymbol{HFI:FWHM:Maps:217GHz}{4.87}
\setsymbol{HFI:FWHM:Maps:353GHz}{4.65}
\setsymbol{HFI:FWHM:Maps:545GHz}{4.72}
\setsymbol{HFI:FWHM:Maps:857GHz}{4.39}

\setsymbol{HFI:beam:ellipticity:Maps:100GHz}{1.15}
\setsymbol{HFI:beam:ellipticity:Maps:143GHz}{1.01}
\setsymbol{HFI:beam:ellipticity:Maps:217GHz}{1.06}
\setsymbol{HFI:beam:ellipticity:Maps:353GHz}{1.05}
\setsymbol{HFI:beam:ellipticity:Maps:545GHz}{1.14}
\setsymbol{HFI:beam:ellipticity:Maps:857GHz}{1.19}

\setsymbol{HFI:FWHM:Mars:100GHz:units}{9\parcm37}
\setsymbol{HFI:FWHM:Mars:143GHz:units}{7\parcm04}
\setsymbol{HFI:FWHM:Mars:217GHz:units}{4\parcm68}
\setsymbol{HFI:FWHM:Mars:353GHz:units}{4\parcm43}
\setsymbol{HFI:FWHM:Mars:545GHz:units}{3\parcm80}
\setsymbol{HFI:FWHM:Mars:857GHz:units}{3\parcm67}

\setsymbol{HFI:FWHM:Mars:100GHz}{9.37}
\setsymbol{HFI:FWHM:Mars:143GHz}{7.04}
\setsymbol{HFI:FWHM:Mars:217GHz}{4.68}
\setsymbol{HFI:FWHM:Mars:353GHz}{4.43}
\setsymbol{HFI:FWHM:Mars:545GHz}{3.80}
\setsymbol{HFI:FWHM:Mars:857GHz}{3.67}

\setsymbol{HFI:beam:ellipticity:Mars:100GHz}{1.18}
\setsymbol{HFI:beam:ellipticity:Mars:143GHz}{1.03}
\setsymbol{HFI:beam:ellipticity:Mars:217GHz}{1.14}
\setsymbol{HFI:beam:ellipticity:Mars:353GHz}{1.09}
\setsymbol{HFI:beam:ellipticity:Mars:545GHz}{1.25}
\setsymbol{HFI:beam:ellipticity:Mars:857GHz}{1.03}

\setsymbol{HFI:CMB:relative:calibration:100GHz}{$\lsim 1\%$}
\setsymbol{HFI:CMB:relative:calibration:143GHz}{$\lsim 1\%$}
\setsymbol{HFI:CMB:relative:calibration:217GHz}{$\lsim 1\%$}
\setsymbol{HFI:CMB:relative:calibration:353GHz}{$\lsim 1\%$}
\setsymbol{HFI:CMB:relative:calibration:545GHz}{}
\setsymbol{HFI:CMB:relative:calibration:857GHz}{}

\setsymbol{HFI:CMB:absolute:calibration:100GHz}{$\lsim 2\%$}
\setsymbol{HFI:CMB:absolute:calibration:143GHz}{$\lsim 2\%$}
\setsymbol{HFI:CMB:absolute:calibration:217GHz}{$\lsim 2\%$}
\setsymbol{HFI:CMB:absolute:calibration:353GHz}{$\lsim 2\%$}
\setsymbol{HFI:CMB:absolute:calibration:545GHz}{}
\setsymbol{HFI:CMB:absolute:calibration:857GHz}{}

\setsymbol{HFI:FIRAS:gain:calibration:accuracy:statistical:100GHz}{}
\setsymbol{HFI:FIRAS:gain:calibration:accuracy:statistical:143GHz}{}
\setsymbol{HFI:FIRAS:gain:calibration:accuracy:statistical:217GHz}{}
\setsymbol{HFI:FIRAS:gain:calibration:accuracy:statistical:353GHz}{2.5\%}
\setsymbol{HFI:FIRAS:gain:calibration:accuracy:statistical:545GHz}{1\%}
\setsymbol{HFI:FIRAS:gain:calibration:accuracy:statistical:857GHz}{0.5\%}

\setsymbol{HFI:FIRAS:gain:calibration:accuracy:systematic:100GHz}{}
\setsymbol{HFI:FIRAS:gain:calibration:accuracy:systematic:143GHz}{}
\setsymbol{HFI:FIRAS:gain:calibration:accuracy:systematic:217GHz}{}
\setsymbol{HFI:FIRAS:gain:calibration:accuracy:systematic:353GHz}{}
\setsymbol{HFI:FIRAS:gain:calibration:accuracy:systematic:545GHz}{7\%}
\setsymbol{HFI:FIRAS:gain:calibration:accuracy:systematic:857GHz}{7\%}

\setsymbol{HFI:FIRAS:zero:point:accuracy:100GHz:units}{0.8\MJysr}
\setsymbol{HFI:FIRAS:zero:point:accuracy:143GHz:units}{}
\setsymbol{HFI:FIRAS:zero:point:accuracy:217GHz:units}{}
\setsymbol{HFI:FIRAS:zero:point:accuracy:353GHz:units}{1.4\MJysr}
\setsymbol{HFI:FIRAS:zero:point:accuracy:545GHz:units}{2.2\MJysr}
\setsymbol{HFI:FIRAS:zero:point:accuracy:857GHz:units}{1.7\MJysr}

\setsymbol{HFI:FIRAS:zero:point:accuracy:100GHz}{0.8}
\setsymbol{HFI:FIRAS:zero:point:accuracy:143GHz}{}
\setsymbol{HFI:FIRAS:zero:point:accuracy:217GHz}{}
\setsymbol{HFI:FIRAS:zero:point:accuracy:353GHz}{1.4}
\setsymbol{HFI:FIRAS:zero:point:accuracy:545GHz}{2.2}
\setsymbol{HFI:FIRAS:zero:point:accuracy:857GHz}{1.7}

\setsymbol{HFI:unit:conversion:100GHz:units}{0.2415\MJysrmK}
\setsymbol{HFI:unit:conversion:143GHz:units}{0.3694\MJysrmK}
\setsymbol{HFI:unit:conversion:217GHz:units}{0.4811\MJysrmK}
\setsymbol{HFI:unit:conversion:353GHz:units}{0.2883\MJysrmK}
\setsymbol{HFI:unit:conversion:545GHz:units}{0.05826\MJysrmK}
\setsymbol{HFI:unit:conversion:857GHz:units}{0.002238\MJysrmK}

\setsymbol{HFI:unit:conversion:100GHz}{0.2415}
\setsymbol{HFI:unit:conversion:143GHz}{0.3694}
\setsymbol{HFI:unit:conversion:217GHz}{0.4811}
\setsymbol{HFI:unit:conversion:353GHz}{0.2883}
\setsymbol{HFI:unit:conversion:545GHz}{0.05826}
\setsymbol{HFI:unit:conversion:857GHz}{0.002238}

\setsymbol{HFI:colour:correction:alpha=-2:V1.01:100GHz}{0.9893}
\setsymbol{HFI:colour:correction:alpha=-2:V1.01:143GHz}{0.9759}
\setsymbol{HFI:colour:correction:alpha=-2:V1.01:217GHz}{1.0007}
\setsymbol{HFI:colour:correction:alpha=-2:V1.01:353GHz}{1.0028}
\setsymbol{HFI:colour:correction:alpha=-2:V1.01:545GHz}{1.0019}
\setsymbol{HFI:colour:correction:alpha=-2:V1.01:857GHz}{0.9889}

\setsymbol{HFI:colour:correction:alpha=0:V1.01:100GHz}{1.0008}
\setsymbol{HFI:colour:correction:alpha=0:V1.01:143GHz}{1.0148}
\setsymbol{HFI:colour:correction:alpha=0:V1.01:217GHz}{0.9909}
\setsymbol{HFI:colour:correction:alpha=0:V1.01:353GHz}{0.9888}
\setsymbol{HFI:colour:correction:alpha=0:V1.01:545GHz}{0.9878}
\setsymbol{HFI:colour:correction:alpha=0:V1.01:857GHz}{1.0014}

%% file: AuthorList_P20_Strings_and_Other_Defects_authors_and_institutes.tex
\author{\small
Planck Collaboration:
P.~A.~R.~Ade\inst{83}
\and
N.~Aghanim\inst{55}
\and
C.~Armitage-Caplan\inst{88}
\and
M.~Arnaud\inst{69}
\and
M.~Ashdown\inst{66, 6}
\and
F.~Atrio-Barandela\inst{18}
\and
J.~Aumont\inst{55}
\and
C.~Baccigalupi\inst{82}
\and
A.~J.~Banday\inst{91, 9}
\and
R.~B.~Barreiro\inst{63}
% ...
}
\institute{\small
APC, AstroParticule et Cosmologie, Universit\'{e} Paris Diderot, CNRS/IN2P3, CEA/lrfu, Observatoire de Paris, Sorbonne Paris Cit\'{e}, 10, rue Alice Domon et L\'{e}onie Duquet, 75205 Paris Cedex 13, France\\
\and
Aalto University Mets\"{a}hovi Radio Observatory, Mets\"{a}hovintie 114, FIN-02540 Kylm\"{a}l\"{a}, Finland\\
\and
African Institute for Mathematical Sciences, 6-8 Melrose Road, Muizenberg, Cape Town, South Africa\\
\and
Agenzia Spaziale Italiana Science Data Center, c/o ESRIN, via Galileo Galilei, Frascati, Italy\\
\and
Agenzia Spaziale Italiana, Viale Liegi 26, Roma, Italy\\
\and
Astrophysics Group, Cavendish Laboratory, University of Cambridge, J J Thomson Avenue, Cambridge CB3 0HE, U.K.\\
\and
Astrophysics \& Cosmology Research Unit, School of Mathematics, Statistics \& Computer Science, University of KwaZulu-Natal, Westville Campus, Private Bag X54001, Durban 4000, South Africa\\
\and
CITA, University of Toronto, 60 St. George St., Toronto, ON M5S 3H8, Canada\\
\and
CNRS, IRAP, 9 Av. colonel Roche, BP 44346, F-31028 Toulouse cedex 4, France\\
\and
California Institute of Technology, Pasadena, California, U.S.A.\\
\and
Centre for Theoretical Cosmology, DAMTP, University of Cambridge, Wilberforce Road, Cambridge CB3 0WA U.K.\\
\and
Centro de Estudios de F\'{i}sica del Cosmos de Arag\'{o}n (CEFCA), Plaza San Juan, 1, planta 2, E-44001, Teruel, Spain\\
\and
Computational Cosmology Center, Lawrence Berkeley National Laboratory, Berkeley, California, U.S.A.\\
\and
Consejo Superior de Investigaciones Cient\'{\i}ficas (CSIC), Madrid, Spain\\
\and
DSM/Irfu/SPP, CEA-Saclay, F-91191 Gif-sur-Yvette Cedex, France\\
\and
DTU Space, National Space Institute, Technical University of Denmark, Elektrovej 327, DK-2800 Kgs. Lyngby, Denmark\\
\and
D\'{e}partement de Physique Th\'{e}orique, Universit\'{e} de Gen\`{e}ve, 24, Quai E. Ansermet,1211 Gen\`{e}ve 4, Switzerland\\
\and
Departamento de F\'{\i}sica Fundamental, Facultad de Ciencias, Universidad de Salamanca, 37008 Salamanca, Spain\\
\and
Departamento de F\'{\i}sica, Universidad de Oviedo, Avda. Calvo Sotelo s/n, Oviedo, Spain\\
\and
Department of Astronomy and Astrophysics, University of Toronto, 50 Saint George Street, Toronto, Ontario, Canada\\
\and
Department of Astrophysics/IMAPP, Radboud University Nijmegen, P.O. Box 9010, 6500 GL Nijmegen, The Netherlands\\
\and
Department of Electrical Engineering and Computer Sciences, University of California, Berkeley, California, U.S.A.\\
\and
Department of Physics \& Astronomy, University of British Columbia, 6224 Agricultural Road, Vancouver, British Columbia, Canada\\
\and
Department of Physics and Astronomy, Dana and David Dornsife College of Letter, Arts and Sciences, University of Southern California, Los Angeles, CA 90089, U.S.A.\\
\and
Department of Physics and Astronomy, University College London, London WC1E 6BT, U.K.\\
\and
Department of Physics, Gustaf H\"{a}llstr\"{o}min katu 2a, University of Helsinki, Helsinki, Finland\\
\and
Department of Physics, Princeton University, Princeton, New Jersey, U.S.A.\\
\and
Department of Physics, University of California, One Shields Avenue, Davis, California, U.S.A.\\
\and
Department of Physics, University of California, Santa Barbara, California, U.S.A.\\
\and
Department of Physics, University of Illinois at Urbana-Champaign, 1110 West Green Street, Urbana, Illinois, U.S.A.\\
\and
Dipartimento di Fisica e Astronomia G. Galilei, Universit\`{a} degli Studi di Padova, via Marzolo 8, 35131 Padova, Italy\\
\and
Dipartimento di Fisica e Scienze della Terra, Universit\`{a} di Ferrara, Via Saragat 1, 44122 Ferrara, Italy\\
\and
Dipartimento di Fisica, Universit\`{a} La Sapienza, P. le A. Moro 2, Roma, Italy\\
\and
Dipartimento di Fisica, Universit\`{a} degli Studi di Milano, Via Celoria, 16, Milano, Italy\\
\and
Dipartimento di Fisica, Universit\`{a} degli Studi di Trieste, via A. Valerio 2, Trieste, Italy\\
\and
Dipartimento di Fisica, Universit\`{a} di Roma Tor Vergata, Via della Ricerca Scientifica, 1, Roma, Italy\\
\and
Discovery Center, Niels Bohr Institute, Blegdamsvej 17, Copenhagen, Denmark\\
\and
Dpto. Astrof\'{i}sica, Universidad de La Laguna (ULL), E-38206 La Laguna, Tenerife, Spain\\
\and
European Space Agency, ESAC, Planck Science Office, Camino bajo del Castillo, s/n, Urbanizaci\'{o}n Villafranca del Castillo, Villanueva de la Ca\~{n}ada, Madrid, Spain\\
\and
European Space Agency, ESTEC, Keplerlaan 1, 2201 AZ Noordwijk, The Netherlands\\
\and
Helsinki Institute of Physics, Gustaf H\"{a}llstr\"{o}min katu 2, University of Helsinki, Helsinki, Finland\\
\and
INAF - Osservatorio Astronomico di Padova, Vicolo dell'Osservatorio 5, Padova, Italy\\
\and
INAF - Osservatorio Astronomico di Roma, via di Frascati 33, Monte Porzio Catone, Italy\\
\and
INAF - Osservatorio Astronomico di Trieste, Via G.B. Tiepolo 11, Trieste, Italy\\
\and
INAF/IASF Bologna, Via Gobetti 101, Bologna, Italy\\
\and
INAF/IASF Milano, Via E. Bassini 15, Milano, Italy\\
\and
INFN, Sezione di Bologna, Via Irnerio 46, I-40126, Bologna, Italy\\
\and
INFN, Sezione di Roma 1, Universit\`{a} di Roma Sapienza, Piazzale Aldo Moro 2, 00185, Roma, Italy\\
\and
IPAG: Institut de Plan\'{e}tologie et d'Astrophysique de Grenoble, Universit\'{e} Joseph Fourier, Grenoble 1 / CNRS-INSU, UMR 5274, Grenoble, F-38041, France\\
\and
IUCAA, Post Bag 4, Ganeshkhind, Pune University Campus, Pune 411 007, India\\
\and
Imperial College London, Astrophysics group, Blackett Laboratory, Prince Consort Road, London, SW7 2AZ, U.K.\\
\and
Infrared Processing and Analysis Center, California Institute of Technology, Pasadena, CA 91125, U.S.A.\\
\and
Institut N\'{e}el, CNRS, Universit\'{e} Joseph Fourier Grenoble I, 25 rue des Martyrs, Grenoble, France\\
\and
Institut Universitaire de France, 103, bd Saint-Michel, 75005, Paris, France\\
\and
Institut d'Astrophysique Spatiale, CNRS (UMR8617) Universit\'{e} Paris-Sud 11, B\^{a}timent 121, Orsay, France\\
\and
Institut d'Astrophysique de Paris, CNRS (UMR7095), 98 bis Boulevard Arago, F-75014, Paris, France\\
\and
Institute for Space Sciences, Bucharest-Magurale, Romania\\
\and
Institute of Astronomy and Astrophysics, Academia Sinica, Taipei, Taiwan\\
\and
Institute of Astronomy, University of Cambridge, Madingley Road, Cambridge CB3 0HA, U.K.\\
\and
Institute of Mathematics and Physics, Centre for Cosmology, Particle Physics and Phenomenology, Louvain University, Louvain-la-Neuve, Belgium\\
\and
Institute of Theoretical Astrophysics, University of Oslo, Blindern, Oslo, Norway\\
\and
Instituto de Astrof\'{\i}sica de Canarias, C/V\'{\i}a L\'{a}ctea s/n, La Laguna, Tenerife, Spain\\
\and
Instituto de F\'{\i}sica de Cantabria (CSIC-Universidad de Cantabria), Avda. de los Castros s/n, Santander, Spain\\
\and
Jet Propulsion Laboratory, California Institute of Technology, 4800 Oak Grove Drive, Pasadena, California, U.S.A.\\
\and
Jodrell Bank Centre for Astrophysics, Alan Turing Building, School of Physics and Astronomy, The University of Manchester, Oxford Road, Manchester, M13 9PL, U.K.\\
\and
Kavli Institute for Cosmology Cambridge, Madingley Road, Cambridge, CB3 0HA, U.K.\\
\and
LAL, Universit\'{e} Paris-Sud, CNRS/IN2P3, Orsay, France\\
\and
LERMA, CNRS, Observatoire de Paris, 61 Avenue de l'Observatoire, Paris, France\\
\and
Laboratoire AIM, IRFU/Service d'Astrophysique - CEA/DSM - CNRS - Universit\'{e} Paris Diderot, B\^{a}t. 709, CEA-Saclay, F-91191 Gif-sur-Yvette Cedex, France\\
\and
Laboratoire Traitement et Communication de l'Information, CNRS (UMR 5141) and T\'{e}l\'{e}com ParisTech, 46 rue Barrault F-75634 Paris Cedex 13, France\\
\and
Laboratoire de Physique Subatomique et de Cosmologie, Universit\'{e} Joseph Fourier Grenoble I, CNRS/IN2P3, Institut National Polytechnique de Grenoble, 53 rue des Martyrs, 38026 Grenoble cedex, France\\
\and
Laboratoire de Physique Th\'{e}orique, Universit\'{e} Paris-Sud 11 \& CNRS, B\^{a}timent 210, 91405 Orsay, France\\
\and
Lawrence Berkeley National Laboratory, Berkeley, California, U.S.A.\\
\and
Max-Planck-Institut f\"{u}r Astrophysik, Karl-Schwarzschild-Str. 1, 85741 Garching, Germany\\
\and
Max-Planck-Institut f\"{u}r Extraterrestrische Physik, Giessenbachstra{\ss}e, 85748 Garching, Germany\\
\and
McGill Physics, Ernest Rutherford Physics Building, McGill University, 3600 rue University, Montr\'{e}al, QC, H3A 2T8, Canada\\
\and
MilliLab, VTT Technical Research Centre of Finland, Tietotie 3, Espoo, Finland\\
\and
Niels Bohr Institute, Blegdamsvej 17, Copenhagen, Denmark\\
\and
Observational Cosmology, Mail Stop 367-17, California Institute of Technology, Pasadena, CA, 91125, U.S.A.\\
\and
Optical Science Laboratory, University College London, Gower Street, London, U.K.\\
\and
SB-ITP-LPPC, EPFL, CH-1015, Lausanne, Switzerland\\
\and
SISSA, Astrophysics Sector, via Bonomea 265, 34136, Trieste, Italy\\
\and
School of Physics and Astronomy, Cardiff University, Queens Buildings, The Parade, Cardiff, CF24 3AA, U.K.\\
\and
School of Physics and Astronomy, University of Nottingham, Nottingham NG7 2RD, U.K.\\
\and
Space Sciences Laboratory, University of California, Berkeley, California, U.S.A.\\
\and
Special Astrophysical Observatory, Russian Academy of Sciences, Nizhnij Arkhyz, Zelenchukskiy region, Karachai-Cherkessian Republic, 369167, Russia\\
\and
Stanford University, Dept of Physics, Varian Physics Bldg, 382 Via Pueblo Mall, Stanford, California, U.S.A.\\
\and
Sub-Department of Astrophysics, University of Oxford, Keble Road, Oxford OX1 3RH, U.K.\\
\and
Theory Division, PH-TH, CERN, CH-1211, Geneva 23, Switzerland\\
\and
UPMC Univ Paris 06, UMR7095, 98 bis Boulevard Arago, F-75014, Paris, France\\
\and
Universit\'{e} de Toulouse, UPS-OMP, IRAP, F-31028 Toulouse cedex 4, France\\
% ...
}